\documentclass[traditabstract,longauth]{aa}
\usepackage{txfonts}
\usepackage{graphicx}
\usepackage{natbib}
\usepackage{color}
\usepackage[colorlinks=true,citecolor=blue]{hyperref}

\newcommand{\be}{\begin{equation}}
\newcommand{\ee}{\end{equation}}
\newcommand{\bp}{\begin{figure}[!ht]}
\newcommand{\ep}{\end{figure}}
\newcommand{\bpm}{\begin{figure*}[!ht]}
\newcommand{\epm}{\end{figure*}}
\newcommand{\reffig}[1]{Fig.\,\ref{fig:#1}}
\newcommand{\refsec}[1]{Sect.\,\ref{sec:#1}}
\newcommand{\reftab}[1]{Table \ref{tab:#1}}

\newcommand{\commander}{{\tt Commander}}
\newcommand{\healpix} {{HEALPix}}
\newcommand{\planck} {\textit{Planck}}
\newcommand{\transpose}{^{\rm T}}

\def\Planck{\textit{Planck}}
\def\arcm{\ifmmode {^{\scriptstyle\prime}}
          \else $^{\scriptstyle\prime}$\fi}

\begin{document}

\author{\small
Planck Collaboration:
P.~A.~R.~Ade\inst{83}
\and
N.~Aghanim\inst{57}
\and
M.~Arnaud\inst{72}
\and
M.~Ashdown\inst{68, 5}
\and
F.~Atrio-Barandela\inst{16}
\and
J.~Aumont\inst{57}
\and
C.~Baccigalupi\inst{82}
\and
A.~Balbi\inst{33}
\and
A.~J.~Banday\inst{88, 7}
\and
R.~B.~Barreiro\inst{64}
% ...
}
\institute{\small
APC, AstroParticule et Cosmologie, Universit\'{e} Paris Diderot, CNRS/IN2P3, CEA/lrfu, Observatoire de Paris, Sorbonne Paris Cit\'{e}, 10, rue Alice Domon et L\'{e}onie Duquet, 75205 Paris Cedex 13, France\\
\and
Aalto University Mets\"{a}hovi Radio Observatory, Mets\"{a}hovintie 114, FIN-02540 Kylm\"{a}l\"{a}, Finland\\
\and
Agenzia Spaziale Italiana Science Data Center, c/o ESRIN, via Galileo Galilei, Frascati, Italy\\
\and
Agenzia Spaziale Italiana, Viale Liegi 26, Roma, Italy\\
\and
Astrophysics Group, Cavendish Laboratory, University of Cambridge, J J Thomson Avenue, Cambridge CB3 0HE, U.K.\\
\and
CITA, University of Toronto, 60 St. George St., Toronto, ON M5S 3H8, Canada\\
\and
CNRS, IRAP, 9 Av. colonel Roche, BP 44346, F-31028 Toulouse cedex 4, France\\
\and
California Institute of Technology, Pasadena, California, U.S.A.\\
\and
Centre of Mathematics for Applications, University of Oslo, Blindern, Oslo, Norway\\
\and
Centro de Estudios de F\'{i}sica del Cosmos de Arag\'{o}n (CEFCA), Plaza San Juan, 1, planta 2, E-44001, Teruel, Spain\\
\and
Computational Cosmology Center, Lawrence Berkeley National Laboratory, Berkeley, California, U.S.A.\\
\and
Consejo Superior de Investigaciones Cient\'{\i}ficas (CSIC), Madrid, Spain\\
\and
DSM/Irfu/SPP, CEA-Saclay, F-91191 Gif-sur-Yvette Cedex, France\\
\and
DTU Space, National Space Institute, Technical University of Denmark, Elektrovej 327, DK-2800 Kgs. Lyngby, Denmark\\
\and
D\'{e}partement de Physique Th\'{e}orique, Universit\'{e} de Gen\`{e}ve, 24, Quai E. Ansermet,1211 Gen\`{e}ve 4, Switzerland\\
\and
Departamento de F\'{\i}sica Fundamental, Facultad de Ciencias, Universidad de Salamanca, 37008 Salamanca, Spain\\
\and
Departamento de F\'{\i}sica, Universidad de Oviedo, Avda. Calvo Sotelo s/n, Oviedo, Spain\\
\and
Department of Astrophysics, IMAPP, Radboud University, P.O. Box 9010, 6500 GL Nijmegen,  The Netherlands\\
\and
Department of Physics \&amp; Astronomy, University of British Columbia, 6224 Agricultural Road, Vancouver, British Columbia, Canada\\
\and
Department of Physics and Astronomy, Dana and David Dornsife College of Letter, Arts and Sciences, University of Southern California, Los Angeles, CA 90089, U.S.A.\\
\and
Department of Physics, Gustaf H\"{a}llstr\"{o}min katu 2a, University of Helsinki, Helsinki, Finland\\
\and
Department of Physics, Princeton University, Princeton, New Jersey, U.S.A.\\
\and
Department of Physics, University of California, Berkeley, California, U.S.A.\\
\and
Department of Physics, University of California, One Shields Avenue, Davis, California, U.S.A.\\
\and
Department of Physics, University of California, Santa Barbara, California, U.S.A.\\
\and
Department of Physics, University of Illinois at Urbana-Champaign, 1110 West Green Street, Urbana, Illinois, U.S.A.\\
\and
Department of Statistics, Purdue University, 250 N. University Street, West Lafayette, Indiana, U.S.A.\\
\and
Dipartimento di Fisica e Astronomia G. Galilei, Universit\`{a} degli Studi di Padova, via Marzolo 8, 35131 Padova, Italy\\
\and
Dipartimento di Fisica, Universit\`{a} La Sapienza, P. le A. Moro 2, Roma, Italy\\
\and
Dipartimento di Fisica, Universit\`{a} degli Studi di Milano, Via Celoria, 16, Milano, Italy\\
\and
Dipartimento di Fisica, Universit\`{a} degli Studi di Trieste, via A. Valerio 2, Trieste, Italy\\
\and
Dipartimento di Fisica, Universit\`{a} di Ferrara, Via Saragat 1, 44122 Ferrara, Italy\\
\and
Dipartimento di Fisica, Universit\`{a} di Roma Tor Vergata, Via della Ricerca Scientifica, 1, Roma, Italy\\
\and
Dipartimento di Matematica, Universit\`{a} di Roma Tor Vergata, Via della Ricerca Scientifica, 1, Roma, Italy\\
\and
Discovery Center, Niels Bohr Institute, Blegdamsvej 17, Copenhagen, Denmark\\
\and
Dpto. Astrof\'{i}sica, Universidad de La Laguna (ULL), E-38206 La Laguna, Tenerife, Spain\\
\and
European Space Agency, ESAC, Planck Science Office, Camino bajo del Castillo, s/n, Urbanizaci\'{o}n Villafranca del Castillo, Villanueva de la Ca\~{n}ada, Madrid, Spain\\
\and
European Space Agency, ESTEC, Keplerlaan 1, 2201 AZ Noordwijk, The Netherlands\\
\and
Haverford College Astronomy Department, 370 Lancaster Avenue, Haverford, Pennsylvania, U.S.A.\\
\and
Helsinki Institute of Physics, Gustaf H\"{a}llstr\"{o}min katu 2, University of Helsinki, Helsinki, Finland\\
\and
INAF - Osservatorio Astrofisico di Catania, Via S. Sofia 78, Catania, Italy\\
\and
INAF - Osservatorio Astronomico di Padova, Vicolo dell'Osservatorio 5, Padova, Italy\\
\and
INAF - Osservatorio Astronomico di Roma, via di Frascati 33, Monte Porzio Catone, Italy\\
\and
INAF - Osservatorio Astronomico di Trieste, Via G.B. Tiepolo 11, Trieste, Italy\\
\and
INAF Istituto di Radioastronomia, Via P. Gobetti 101, 40129 Bologna, Italy\\
\and
INAF/IASF Bologna, Via Gobetti 101, Bologna, Italy\\
\and
INAF/IASF Milano, Via E. Bassini 15, Milano, Italy\\
\and
INFN, Sezione di Roma 1, Universit`{a} di Roma Sapienza, Piazzale Aldo Moro 2, 00185, Roma, Italy\\
\and
INRIA, Laboratoire de Recherche en Informatique, Universit\'{e} Paris-Sud 11, B\^{a}timent 490, 91405 Orsay Cedex, France\\
\and
IPAG: Institut de Plan\'{e}tologie et d'Astrophysique de Grenoble, Universit\'{e} Joseph Fourier, Grenoble 1 / CNRS-INSU, UMR 5274, Grenoble, F-38041, France\\
\and
ISDC Data Centre for Astrophysics, University of Geneva, ch. d'Ecogia 16, Versoix, Switzerland\\
\and
IUCAA, Post Bag 4, Ganeshkhind, Pune University Campus, Pune 411 007, India\\
\and
Imperial College London, Astrophysics group, Blackett Laboratory, Prince Consort Road, London, SW7 2AZ, U.K.\\
\and
Infrared Processing and Analysis Center, California Institute of Technology, Pasadena, CA 91125, U.S.A.\\
\and
Institut N\'{e}el, CNRS, Universit\'{e} Joseph Fourier Grenoble I, 25 rue des Martyrs, Grenoble, France\\
\and
Institut Universitaire de France, 103, bd Saint-Michel, 75005, Paris, France\\
\and
Institut d'Astrophysique Spatiale, CNRS (UMR8617) Universit\'{e} Paris-Sud 11, B\^{a}timent 121, Orsay, France\\
\and
Institut d'Astrophysique de Paris, CNRS (UMR7095), 98 bis Boulevard Arago, F-75014, Paris, France\\
\and
Institute for Space Sciences, Bucharest-Magurale, Romania\\
\and
Institute of Astronomy and Astrophysics, Academia Sinica, Taipei, Taiwan\\
\and
Institute of Astronomy, University of Cambridge, Madingley Road, Cambridge CB3 0HA, U.K.\\
\and
Institute of Theoretical Astrophysics, University of Oslo, Blindern, Oslo, Norway\\
\and
Instituto de Astrof\'{\i}sica de Canarias, C/V\'{\i}a L\'{a}ctea s/n, La Laguna, Tenerife, Spain\\
\and
Instituto de F\'{\i}sica de Cantabria (CSIC-Universidad de Cantabria), Avda. de los Castros s/n, Santander, Spain\\
\and
Istituto di Fisica del Plasma, CNR-ENEA-EURATOM Association, Via R. Cozzi 53, Milano, Italy\\
\and
Jet Propulsion Laboratory, California Institute of Technology, 4800 Oak Grove Drive, Pasadena, California, U.S.A.\\
\and
Jodrell Bank Centre for Astrophysics, Alan Turing Building, School of Physics and Astronomy, The University of Manchester, Oxford Road, Manchester, M13 9PL, U.K.\\
\and
Kavli Institute for Cosmology Cambridge, Madingley Road, Cambridge, CB3 0HA, U.K.\\
\and
Kavli Institute for Theoretical Physics, University of California, Santa Barbara Kohn Hall, Santa Barbara, CA 93106, U.S.A.\\
\and
LAL, Universit\'{e} Paris-Sud, CNRS/IN2P3, Orsay, France\\
\and
LERMA, CNRS, Observatoire de Paris, 61 Avenue de l'Observatoire, Paris, France\\
\and
Laboratoire AIM, IRFU/Service d'Astrophysique - CEA/DSM - CNRS - Universit\'{e} Paris Diderot, B\^{a}t. 709, CEA-Saclay, F-91191 Gif-sur-Yvette Cedex, France\\
\and
Laboratoire Traitement et Communication de l'Information, CNRS (UMR 5141) and T\'{e}l\'{e}com ParisTech, 46 rue Barrault F-75634 Paris Cedex 13, France\\
\and
Laboratoire de Physique Subatomique et de Cosmologie, Universit\'{e} Joseph Fourier Grenoble I, CNRS/IN2P3, Institut National Polytechnique de Grenoble, 53 rue des Martyrs, 38026 Grenoble cedex, France\\
\and
Laboratoire de Physique Th\'{e}orique, Universit\'{e} Paris-Sud 11 \&amp; CNRS, B\^{a}timent 210, 91405 Orsay, France\\
\and
Lawrence Berkeley National Laboratory, Berkeley, California, U.S.A.\\
\and
Max-Planck-Institut f\"{u}r Astrophysik, Karl-Schwarzschild-Str. 1, 85741 Garching, Germany\\
\and
National University of Ireland, Department of Experimental Physics, Maynooth, Co. Kildare, Ireland\\
\and
Niels Bohr Institute, Blegdamsvej 17, Copenhagen, Denmark\\
\and
Observational Cosmology, Mail Stop 367-17, California Institute of Technology, Pasadena, CA, 91125, U.S.A.\\
\and
Optical Science Laboratory, University College London, Gower Street, London, U.K.\\
\and
SISSA, Astrophysics Sector, via Bonomea 265, 34136, Trieste, Italy\\
\and
School of Physics and Astronomy, Cardiff University, Queens Buildings, The Parade, Cardiff, CF24 3AA, U.K.\\
\and
Space Sciences Laboratory, University of California, Berkeley, California, U.S.A.\\
\and
Stanford University, Dept of Physics, Varian Physics Bldg, 382 Via Pueblo Mall, Stanford, California, U.S.A.\\
\and
UPMC Univ Paris 06, UMR7095, 98 bis Boulevard Arago, F-75014, Paris, France\\
\and
Universit\"{a}t Heidelberg, Institut f\"{u}r Theoretische Astrophysik, Albert-\"{U}berle-Str. 2, 69120, Heidelberg, Germany\\
\and
Universit\'{e} de Toulouse, UPS-OMP, IRAP, F-31028 Toulouse cedex 4, France\\
% ...
}

\title{\planck\ intermediate results. IX. Detection of the Galactic haze with \Planck}

\authorrunning{Planck Collaboration}
\titlerunning{Detection of the Galactic haze with \Planck}

\abstract { 
  Using precise full-sky observations from \textit{Planck}, and applying
  several methods of component separation, we identify and
  characterize the emission from the Galactic ``haze'' at microwave
  wavelengths.  The  haze is a distinct component of
  diffuse Galactic emission, roughly centered on the Galactic
  centre, and extends to $|b|\sim35\degr$ in Galactic latitude and
  $|l|\sim15\degr$ in longitude.  By combining the \textit{Planck}
  data with observations from the \textit{Wilkinson Microwave Anisotropy
    Probe} we are able to determine the spectrum of this emission to
  high accuracy, unhindered by the large systematic biases present in
  previous analyses.  The derived spectrum is consistent with
  power-law emission with a spectral index of $-2.55 \pm 0.05$, thus
  excluding free-free emission as the source  and instead
  favouring hard-spectrum synchrotron radiation from an electron population with
  a spectrum (number density per energy) $dN/dE \propto E^{-2.1}$.  At
  Galactic latitudes $|b|<30\degr$, the microwave haze morphology is
  consistent with that of the \textit{Fermi} gamma-ray ``haze'' or ``bubbles,''
  indicating that we have a multi-wavelength view of a distinct
  component of our Galaxy.  Given both the very hard spectrum and the
  extended nature of the emission, it is highly unlikely that the haze
  electrons result from supernova shocks in the Galactic
  disk. Instead, a new mechanism for cosmic-ray acceleration in the
  centre of our Galaxy is implied.
}
 \maketitle

\section{Introduction}
\label{sec:introduction}
The initial data release from the \textit{Wilkinson Microwave Anisotropy
Probe} (\textit{WMAP}) revolutionised our understanding of both cosmology
\citep{spergel03} and the physical processes at work in the
interstellar medium (ISM) of our own Galaxy \citep{bennett03b}.  Some
of the processes observed were expected, such as
the thermal emission from dust grains, free-free emission (or thermal
bremsstrahlung) from electron/ion scattering, and synchrotron emission
due to shock-accelerated electrons interacting with the Galactic
magnetic field.  Others, such as the anomalous microwave emission now
identified as spinning dust emission from rapidly rotating tiny dust
grains
\citep{draine98a,draine98b,deoliveiracosta02,finkbeiner04c,hinshaw07,boughn07,dobler08b,dobler09},
were more surprising. But perhaps most mysterious was a ``haze'' of
emission discovered by \citet{finkbeiner04a} that was centred on the
Galactic centre (GC), appeared roughly spherically symmetric in
profile, fell off roughly as the inverse distance from the GC, and was
of unknown origin.  This haze was originally characterised
as free-free emission by \citet{finkbeiner04a} due to its apparently very
hard spectrum, although it was not appreciated at the time how
significant the systematic uncertainty in the measured spectrum was.

\defcitealias{dobler08a}{DF08}

An analysis of the 3-year \textit{WMAP} data by \citet[hereafter
DF08]{dobler08a}
identified a source of systematic uncertainty in the determination of
the haze spectrum that remains the key to determining the origin of
the emission.  This uncertainty is due to residual foregrounds
contaminating the cosmic microwave background (CMB) radiation estimate
used in the analysis, and arises as a consequence of chance
morphological correlations between the CMB and the haze itself.
Nevertheless, the spectrum was found to be both significantly softer
than free-free emission, and also significantly harder than
the synchrotron emission observed elsewhere in the Galaxy as traced by
the low-frequency synchrotron measurements of \citet{haslam82}
\citep[see also][]{reich88,davies96,kogut07,strong11,kogut12}.  Finally, it
was noted that this systematic uncertainty could be almost completely
eliminated with data from the \textit{Planck}\footnote{\Planck\ (\url{http://www.esa.int/Planck}) is a project of the European Space Agency (ESA) with instruments provided by two scientific consortia funded by ESA member states (in particular the lead countries France and Italy), with contributions from NASA (USA) and telescope reflectors provided by a collaboration between ESA and a scientific consortium led and funded by Denmark.} mission, which would
produce estimates of the CMB signal that were significantly less
contaminated by Galactic foregrounds.

The synchrotron nature of the microwave haze was substantially
supported by the discovery of a gamma-ray counterpart to this emission
by \citet{dobler10} using data from the \textit{Fermi Gamma-Ray Space
Telescope}.  These observations were consistent with an inverse Compton (IC)
signal generated by electrons with the same spectrum and amplitude
as would yield the microwave haze at \textit{WMAP} wavelengths.  Further work
by \citet{su10} showed that the \textit{Fermi} haze appeared to have
sharp edges and it was renamed the ``\textit{Fermi} bubbles.''
Subsequently, there has been significant theoretical interest in determining the
origin of the very hard spectrum of progenitor electrons.  Suggestions
include enhanced supernova rates \citep{biermann10}, a Galactic wind
\citep{crocker11}, a jet generated by accretion onto the central black hole 
\citep{guo11a,guo11b}, and co-annihilation of dark
matter (DM) particles in the Galactic halo
\citep{finkbeiner04b,hooper07,lin10,dobler11a}. However, while each of
these scenarios can reproduce some of the properties of the
haze/bubbles well, none can completely match all of the observed
characteristics \citep{dobler12}.

Moreover, despite the significant observational evidence, there have  been 
suggestions in the literature that the microwave haze is either 
an artefact of the analysis procedure \citep{mertsch10} or not synchrotron
emission \citep{gold11}.  The former conclusion was initially
supported by 
alternative analyses of the \textit{WMAP} data 
that found no evidence of the haze \citep{eriksen06, dickinson09}.
However, more recently \citet{pietrobon11} showed that these analyses, while
extremely effective at cleaning the CMB of foregrounds and identifying
likely contaminants of a known morphology (e.g., a low-level residual
cosmological dipole), 
typically cannot separate the haze emission from a low-frequency 
combination of free-free, spinning dust, and softer
synchrotron radiation.  The argument of \citet{gold11} that the microwave haze is
not synchrotron emission
was based on the lack of detection
of a polarised component.  This criticism was addressed
by \citet{dobler12} who showed that, even if the emission is not
depolarised by turbulence in the magnetic field, such a polarised
signal is not likely to be seen with \textit{WMAP} given the noise in the data.

With the \textit{Planck} data, we now have the ability not only to
provide evidence for the existence of the microwave haze with an
independent experiment, but also to eliminate the
uncertainty in the spectrum of the emission which has hindered both
observational and theoretical studies for nearly a decade.  In
\refsec{data} we describe the \textit{Planck} data as well as some external
templates we use in our analysis.  In \refsec{methods} we describe the
two most effective component separation techniques for studying the
haze emission in temperature. In \refsec{results} we discuss our results on the morphology
and spectrum of the haze, before summarising in
\refsec{summary}.

\section{\textit{Planck} data and templates}
\label{sec:data}
\planck\ \citep{tauber2010a, Planck2011-1.1} is the third generation
space mission to measure the anisotropy of the cosmic microwave
background (CMB).  It observes the sky in nine frequency bands
covering 30--857\,GHz with high sensitivity and angular resolution
from 31\arcm\ to 5\arcm.  The Low Frequency Instrument (LFI;
\citealt{Mandolesi2010, Bersanelli2010, Planck2011-1.4}) covers the
30, 44, and 70\,GHz bands with amplifiers cooled to 20\,\hbox{K}.  The
High Frequency Instrument (HFI; \citealt{Lamarre2010, Planck2011-1.5})
covers the 100, 143, 217, 353, 545, and 857\,GHz bands with bolometers
cooled to 0.1\,\hbox{K}.  Polarisation is measured in all but the
highest two bands \citep{Leahy2010, Rosset2010}.  A combination of
radiative cooling and three mechanical coolers produces the
temperatures needed for the detectors and optics
\citep{Planck2011-1.3}.  Two data processing centres (DPCs) check and
calibrate the data and make maps of the sky \citep{Planck2011-1.7,
  Planck2011-1.6}.  \planck's sensitivity, angular resolution, and
frequency coverage make it a powerful instrument for galactic and
extragalactic astrophysics as well as cosmology.  Early astrophysics
results are given in Planck Collaboration VIII--XXVI 2011, based on
data taken between 13~August 2009 and 7~June 2010.  Intermediate
astrophysics results are now being presented in a series of papers
based on data taken between 13~August 2009 and 27~November 2010.

We take both the \textit{WMAP} and \planck\ bandpasses into account
when defining our central frequencies.  However, throughout we refer
to the bands by the conventional labels of 23, 33, 41, 61, and 94\,GHz
for \textit{WMAP} and 30, 44, 70, 100, 143, 217, 353, 545, and
857\,GHz for \planck; the central frequencies are 22.8, 33.2, 41.0,
61.4, and 94.0\,GHz, and 28.5, 44.1, 70.3, 100.0, 143.0, 217.0, 353.0,
545.0, and 857.0\,GHz respectively.  In each case, the central
frequency represents the convolution of the bandpass response with a
CMB spectrum and so corresponds to the effective frequency for
emission with that spectrum.  For emission with different spectra, the
effective frequency is slightly shifted, but the effects are at the
few percent level and do not significantly affect our conclusions.

Our analysis also requires the use of external templates to
morphologically trace emission mechanisms within the \textit{Planck}
data.  All the data are available in the \healpix\ \footnote{
see http://healpix.jpl.nasa.gov} scheme
\citep{gorski05}.  In each case, we use maps smoothed to 1$\degr$
angular resolution.

\defcitealias{finkbeiner99}{FDS99}

\paragraph{Thermal and spinning dust} For a template of the combined
thermal and spinning dust emission, we use the 100\,$\mu$m all-sky map
from \citet{schlegel98} evaluated at the appropriate
\textit{Planck} and \textit{WMAP}  frequencies using Model~8
from \citet[][FDS99]{finkbeiner99}.  This is a sufficiently good
estimate of the thermal emission for our purposes, although it is
important to note that the morphological correlation between thermal
and spinning dust is not well known.

\paragraph{Free-free} The free-free template adopted in our analysis
is the H$\alpha$ map assembled by \citet{finkbeiner03}\footnote{
Our specific choice of the \citet{finkbeiner03} H$\alpha$ template does
not have a strong impact on results.  We have repeated our analysis
using the \citet{dickinson03} H$\alpha$ map and find differences at the
few percent level that are not spatially correlated with haze
emission.}  from three surveys: the Wisconsin H$\alpha$ Mapper
\citep{haffner03}, the Southern H$\alpha$ Sky Survey Atlas
\citep{gaustad01}, and the Virginia Tech Spectral-Line Survey
\citep{dennison98}.  The map is corrected for line-of-sight dust
absorption assuming uniform mixing between gas and dust, although we
mask some regions based on the predicted total dust extinction where
the correction to the H$\alpha$ emission is deemed unreliable.

\paragraph{Soft Synchrotron} Since synchrotron intensity rises with
decreasing frequency, the 408\,MHz full-sky radio continuum map
\citep{haslam82} provides a reasonable tracer of the
soft synchrotron emission.  While there is a very small contribution
from free-free emission to the observed intensity, particularly in
the Galactic plane, the bulk of the emission traces synchrotron
radiation from supernova shock-accelerated electrons that have had
sufficient time to diffuse from their source.  In addition, as pointed
out by \citet{dobler12}, the propagation length for cosmic-ray
electrons in the disk is energy-dependent and therefore the 408\,MHz
map (which is dominated by synchrotron emission from lower energy
electrons compared to the situation at 20--100\,GHz) will be more
spatially extended than the synchrotron at \planck\ frequencies
\citep[see][]{mertsch10}.  This can result in a disk-like residual
when using the 408\,MHz map as a tracer of synchrotron at higher
frequencies that could be confused with the haze emission.  We
use an elliptical Gaussian disk template ($\sigma_{l}=20\degr$ and
$\sigma_{b}=5\degr$) for this residual, though in practice this
results in only a very small correction to our results, which use a
larger mask than \citet{dobler12} (see below).

\paragraph{The Haze} Although a measurement of the precise morphology of
the microwave haze is to be determined, an estimate of the morphology
is necessary to reduce bias in template fits for the following reason:
when using templates to separate foregrounds, the amplitudes of the
other templates may be biased to compensate for the haze emission
present in the data unless an appropriate haze template is used to
approximate the emission.  Following \citet{dobler12}, we use an
elliptical Gaussian template with $\sigma_{l}=15\degr$ and
$\sigma_{b}=25\degr$.  Note that a map of the \textit{Fermi} gamma-ray
haze/bubbles cannot be used to trace the emission for two reasons.
First, as pointed out by \citet{dobler11a}, the morphology of the
gamma-ray emission is uncertain at low latitudes.  Second, the
synchrotron morphology depends sensitively on the magnetic field while
the gamma-ray morphology depends on the interstellar radiation
field. Therefore, while the same cosmic-ray population is clearly
responsible for both, the detailed morphologies are not
identical.\footnote{We have performed our fits using the uniform
  ``bubbles'' template given in \cite{su10} and the morphology of the
  haze excess (see \refsec{results}) is not significantly changed.}

\paragraph{Mask} As noted above, the effect of dust extinction requires
careful treatment of the H$\alpha$ map when using it as a tracer of
free-free emission.  Therefore, we mask out all regions where dust
extinction at H$\alpha$ wavelengths is greater than 1\,mag.  We
also mask out all point sources in the \textit{WMAP} and \textit{Planck} ERCSC
(30--143\,GHz) catalogs.  Several larger-scale features where our
templates are likely to fail are also masked: the LMC, SMC, M31,
Orion--Barnard's Loop, NGC\,5128, and $\zeta$~Oph.  Finally, since the
H$\alpha$ to free-free ratio is a function of gas temperature, we
mask pixels with H$\alpha$ intensity greater than 10\,rayleigh to
minimise the bias due to strong spatial fluctuations in gas
temperatures.  This mask covers 32\% of the sky and is shown in
\reffig{templates}.

\section{Component separation methods}
\label{sec:methods}
In this paper, we apply two methods for separating the Galactic
emission components in the \textit{Planck} data.  The first one, used in
the original \textit{WMAP} haze analyses, is a simple regression technique in
which the templates described in the previous section are fit directly
to the data.  This ``template fitting'' method is relatively simple to
implement and its results are easy to interpret.  Furthermore, the
noise characteristics are well understood and additional components
not represented by the templates are readily identifiable in residual
maps.  The second technique, a powerful 
power-spectrum estimation and component-separation method based on Gibbs
sampling, uses a Bayesian approach and combines pixel-by-pixel
spectral fits with template amplitudes.  One of the significant
advantages of this approach is that, rather than assuming an estimate
for the CMB anisotropy, a CMB map is generated via joint sampling of
the foreground parameters and $C_{\ell}$s of cosmological
anisotropies; this should reduce the bias in the inferred
foreground spectra.

\subsection{Template fitting}
\label{sec:tempfit}
The rationale behind the simple template fitting technique 
is that there are only a few physical mechanisms in the interstellar
medium that generate emission at microwave wavelengths, and these
emission mechanisms are morphologically traced by maps at other
frequencies at which they dominate.  We follow the linear regression
formalism of \citet{finkbeiner04a}, \citet{dobler08a}, and \citet{dobler12} and solve the
relation
\be
  \vec{d}_{\nu} = \vec{a}_{\nu}\cdot{\bf P},
\ee
where $\vec{d}_{\nu}$ is a data map at frequency $\nu$, ${\bf P}$
is a matrix of the templates defined in \refsec{data}, and 
$\vec{a}_{\nu}$ is the vector of scaling amplitudes for this set of
templates.  The least-squares solution to this equation is
\be
  \vec{a}_{\nu} = ({\bf P}\transpose{\bf N}_{\nu}^{-1}{\bf P})^{-1}({\bf
    P}\transpose{\bf N}_{\nu}^{-1}\vec{d}_{\nu}),
\label{eq:tempfit}
\ee
where ${\bf N}_{\nu}$ is the noise covariance matrix  at frequency $\nu$.  In
practice, for our template fits we use the mean noise per band (i.e.,
we set ${\bf N}_{\nu} = \langle{\bf N}_{\nu}\rangle$ for all pixels),
which is appropriate in the limit where the dominant uncertainty is
how well the templates trace the foregrounds, as is the case here.  To
the extent that the templates morphologically match the actual
foregrounds, the solutions $a^i_{\nu}$ for template $i$ as a function
of frequency represent a reasonable estimate of the spectrum over the
fitted pixels.

There are two important features of this approach to template fitting
that must be addressed.  First, there is an implicit assumption that
the spectrum of a given template-correlated emission mechanism does
not vary across the region of interest, and second, an estimate for
the CMB must be pre-subtracted from the data.  The former can be
validated by inspecting a map of the residuals which can reveal where
this assumption fails, and as a consequence of which the sky can
easily be subdivided into regions that can be fitted independently.
The latter involves the complication that no CMB estimate is
completely clean of the foregrounds to be measured, which thereore
introduces a bias (with the same spectrum as the CMB) in the inferred
foreground spectra.  As shown by \citetalias{dobler08a}, this bias becomes increasingly
large with frequency and renders an exact measurement of the haze
spectrum impossible with \textit{WMAP} alone.  This ``CMB bias'' is the
dominant source of uncertainty in all foreground analyses.  However,
\citetalias{dobler08a} also pointed out that, because the haze spectrum falls with
frequency, the high-frequency data from \textit{Planck} can be used to
generate a CMB estimate that is nearly completely free from haze
emission. Thus, pre-subtraction of this estimate should result in an
essentially unbiased estimate of the haze spectrum.  The CMB estimate
that we use consists of a ``\textit{Planck} HFI internal linear
combination'' (PILC) map, formed from a minimum-variance linear
combination of the \textit{Planck} HFI 143--545\,GHz data after
pre-subtraction of the thermal dust model of \citetalias{finkbeiner99} at each
frequency.\footnote{ Pre-subtracting the \citetalias{finkbeiner99} prediction for the
  thermal dust is not meant to provide a perfect model for the
  thermal dust, but rather a reasonable model.  The goal is to
  minimise variance in the PILC and it is more effective to do so
  by pre-subtracting the
  dust model.  This allows the fit to manage the CO contamination
  present at various HFI frequency channels
  more effectively (although there is still some leakage however, see
  \refsec{planckresults}).  We have tested a PILC which does not
  subtract the thermal dust and the morphology and amplitude of the
  recovered haze signal are similar.} 
Defining $\vec{p}_{\nu}$ and
$\vec{t}_{\nu}$ to be the \textit{Planck} maps and \citetalias{finkbeiner99} prediction
(respectively) at frequency $\nu$, the PILC in $\Delta T_{\rm CMB}$ is
given by \be
  \begin{array}{ccl}
    {\rm PILC} & = & 1.39 \times (\vec{p}_{143} - \vec{t}_{143}) \ - \
    0.36 \times (\vec{p}_{217} - \vec{t}_{217}) \\ & & \ - \
    0.025 \times (\vec{p}_{353} - \vec{t}_{353}) + 0.0013 \times (\vec{p}_{545} - \vec{t}_{545}).
  \end{array}
\ee
The weights are determined by minimising the the variance over
unmasked pixels of the PILC while maintaining a unity response to the
CMB spectrum.

Although no constraint is made on the spectral dependence of the
template coefficients in Eq.~\ref{eq:tempfit}, the fit does
assume that the spectrum is constant across the sky.  While this
assumption is actually quite good outside our mask (as we show
below), it is known to be insufficient in detail.  As such, in
addition to full (unmasked) sky fits, we also perform template fits on
smaller sky regions and combine the results to form a full composite
map.  The subdivisions are defined by hand to separate the sky into
regions with particularly large residuals in a full-sky fit and are
listed in \reftab{tempreg}.

\begin{table}[t]
 \caption{Regions used for the multi-region (RG) template fits.}
 \label{tab:tempreg}
 \centering
    \begin{tabular}{lll}
      \noalign{\hrule\vskip2pt\hrule\vskip 4pt}
      Region & \multicolumn{2}{c}{Sky Coverage} \\
      \noalign{\vskip 3pt\hrule\vskip 5pt}
      1 & $-125\degr \leq l < -104\degr$ &        $-30\degr \leq b <  0\degr$ \\
      2 & $-104\degr \leq l <  -80\degr$ &       $-30\degr \leq b <  0\degr$ \\
      3 & $-125\degr \leq l < -104\degr$ &       $  0\degr \leq b < 30\degr$ \\
      4 & $-104\degr \leq l <  -80\degr$ &       $  0\degr \leq b < 30\degr$ \\
      5 & $ -37\degr \leq l <   42\degr$ &       $  0\degr \leq b < 90\degr$ \\
      6 & $ -80\degr \leq l <  -25\degr$ &       $-30\degr \leq b < 0\degr$ \\
      7 & $  70\degr \leq l <  180\degr$ &       $-90\degr \leq b < 0\degr$ \\
      8 & $ 12\degr \leq l <   70\degr$ &       $-90\degr \leq b < 0\degr$ \\
      9 & \multicolumn{2}{l}{Unmasked pixels outside regions 1--8 and $b \leq 0$} \\
      10 & \multicolumn{2}{l}{Unmasked pixels outside regions 1--8 and  $b > 0$} \\
      \noalign{\vskip3pt\hrule}
    \end{tabular}
 \end{table}

\subsection{Gibbs sampling: Commander}
\label{sec:gibbssamp}
An alternative method for minimising the CMB bias is to generate a CMB
estimate from the data while simultaneously solving for the parameters
of a Galactic foreground model. Within the Bayesian framework it is
possible to set stronger priors on the CMB parameterisation (i.e.,
$C_{\ell}$s), taking advantage not only of the frequency spectrum of
the CMB (a blackbody), but also of the angular power spectrum of the
fluctuations.  Even for relatively simple foreground models, the
dimensionality of parameter space is quite large so uniform
sampling on a grid is not feasible.

\citet{Jewell2004ApJ609} and \citet{Wandelt2004PhRvD70} first
discussed the application of Gibbs sampling algorithms (a variant of
MCMC sampling) in this context. These algorithms have been further improved 
\citep{Eriksen2004ApJS155, ODwyer2004ApJ617L, eriksen07,
Chu2005PhRvD71, Jewell2009ApJ697, Rudjord2009ApJ692, Larson2007ApJ656}
and packaged into the \commander\ code.

Gibbs sampling is particularly suitable for component
separation since it samples from the conditional distribution along
perpendicular directions in parameter space, updating the distribution
with each sample.  This approach has been advocated by
\citet{eriksen07,Eriksen2008ApJ672L} and \citet{dickinson09} and has been
applied recently to the \textit{WMAP} 7-year data by \citet{pietrobon11}. A
detailed description of the algorithm and its validation on simulated
data is provided by \citet[][and references therein]{eriksen08}.

The outputs of the sampling are a map-based CMB estimate and the
parameters of a foreground model, which can either be template-based,
pixel-based, or a combination of the two.  We perform the analysis at
\healpix\ resolution $N_{\rm side}=128$.  The choice of the
foreground model is limited by the number of frequency channels
observed since it sets the number of constraints on the model when
fitting spectra for each pixel. We separate our results in the
following section into two categories, fits using \planck\ data only
and fits using \planck\ data plus ancillary data sets.

For the \planck-only fits, our model consists of a single power law $T
\propto \nu^{\,\beta_{\rm S}}$ describing the effective low-frequency
emission (with a prior on spectral index, $\beta_{\rm
  S}=-3.05\pm0.3$), a grey-body for the thermal dust emission that
dominates at high frequencies (with a temperature and emissivity prior
given by the results of \citealt
{Planck_collaboration:2011A&A...536A..19P}, where mean values of
$T_{\rm D} \simeq 18$\,K and $\epsilon_{\rm D} = 1.8$ were measured),
and a CO spectrum.
The CO spectrum is assumed constant across the sky and normalised to
100\,GHz. The relative strength of the $J$=2$\rightarrow$1
($\sim217$\,GHz) and $J$=3$\rightarrow$2 ($\sim353$\,GHz) transition
lines with respect to the $J$=1$\rightarrow$0 transition were computed
by taking into account the specifications of the HFI detectors and
calibrated by means of the available survey \citep{Dame2001ApJ}.  The
relative ratios in the 100, 217, and 353\,GHz bands are 1.0, 0.35, and
0.12 respectively.  We checked the robustness of the result against a
plausible variation of the line ratios of $\sim10\%$.  (A more
detailed discussion of the CO analysis that we performed can be found
in \citealt{Planck_collaboration:2011A&A...536A..19P}).  We normalise
the thermal dust component at 353\,GHz and the low-frequency power law
at 33\,GHz.  Hence, we solve for two spectral indices together with
the corresponding amplitudes as well as a CO amplitude, with the dust
temperature fixed at a value of 18\,K. The current
\commander\ implementation allows for the determination of residual
monopole and dipole contributions, as may result from the calibration
and map-making procedures.  This fit is referred to as CMD1
throughout.  It is interesting to note that, given the noise in the
data, this highly over-simplified model is sufficient to describe the
total Galactic emission (see \refsec{planckresults}).  However, it is
well established that the low-frequency emission actually consists of
several components.  Following \citet{pietrobon11}, our procedure for
separating these components is to perform a template fit as specified
in Eq.~\ref{eq:tempfit} on the \commander\ solution for the
low-frequency amplitude (i.e., replacing $\vec{d}_{\nu}$ with the
low-frequency amplitude map).  \citet{pietrobon11} showed that
applying this ``post-processing'' template regression procedure is
effective in extracting the haze from the \commander\ solution.

The addition of the \textit{WMAP} channels allows us to refine the foreground
model further, separating the multiple contributions in the frequency
range 23--70\,GHz.  Moreover, the inclusion of the 408\,MHz data
improves the characterisation of the synchrotron component and will
allow us to investigate the spatial variations of its spectral index
(see \refsec{planckwmapresults}). The \commander\ fit, CMD2,
is then based on 14 frequency maps (eight \planck\ channels from 30 to 353\,GHz, 
five from \textit{WMAP}, and Haslam 408\,MHz), and allows a modification of the
foreground model to encompass
two low-frequency power-law components -- one soft component with a
fixed spectral index $\beta_{\rm S} = -3.05$ to describe the soft
synchrotron emission\footnote{This value represents the spectral index
of the large Loop~I feature that is a prominent supernova remnant
visible at both 408\,MHz and microwave frequencies in the northern
Galactic hemisphere.  We have repeated our analysis varying this index
by $\delta\beta=0.1$ and find no significant difference in our
results.}  and one with a spectral index $\beta_{\rm H}$ with prior
$\beta_{\rm H}=-2.15 \pm 0.3$ to capture both the hard synchrotron haze and
the free-free emission.  With this model, the low-frequency part of
the spectrum is more easily resolved into physically meaningful
components.

In addition, we parameterise a joint thermal and spinning-dust model
by
\be
 D_{\rm jd}(\nu) = \left(\frac{\nu}{\nu_0}\right)^{1+\epsilon} \frac{B(\nu,
T)}{B(\nu_0, T)} + {\rm e}^\alpha {\rm e}^{-[(\nu-\nu_1)/b]^{2}/2}.  
\ee 
This is the sum of a grey-body spectrum for the thermal dust, and a
Gaussian profile to mimic the spinning dust SED.  The latter is a
purely phenomenological model selected on the basis of its
straightforward numerical implementation.  However, we have
established its effectiveness in describing well-known spinning dust
regions in the Gould Belt (\planck\ Intermediate Paper, {in
preparation}). The thermal dust pivot frequency $\nu_0$ is set to
545\,GHz and the spinning dust peak frequency $\nu_1$ to 20\,GHz.  The
remaining parameters (the amplitude of the joint spectrum, the
relative amplitude of the spinning dust contribution, and the width of
the spinning dust bump) are constrained by the Gibbs sampling
procedure.  As before we also adopt a spectrum for the CO emission.

\section{Results}
\label{sec:results}

In what follows, we perform four different types of haze
extraction:
\begin{enumerate}
  \item A masked full-sky (FS) template fit for each input frequency
    band. 

  \item Template fits over subsections of the sky (RG) that are
  combined to give a full-sky haze map for each input frequency
  band.

  \item A \commander\ fit (CMD1) with a simple two-component
  foreground model, using \planck\ 30--353\,GHz data.

  \item A comprehensive \commander\ fit (CMD2) including thermal and
  spinning dust models, a soft power-law component, and a hard power-law 
  component, using \planck\ 30--545\,GHz, \textit{WMAP} 23--94\,GHz, and Haslam
  408\,MHz data sets.
\end{enumerate}

We first discuss our results from the template fitting and Gibbs
sampling analyses derived from the \planck\ data alone, then proceed
to include external data sets in the analysis.  A direct comparison of
the results between the template fits and \commander\ haze extraction
methods boosts confidence that, not only are components being
appropriately separated, but the spectrum is relatively free from
bias.

\subsection{\planck-only results}
\label{sec:planckresults}
\subsubsection{Template fitting}
\label{sec:plancktemp}

\bpm
  \centerline{
    \includegraphics[width=0.98\textwidth]{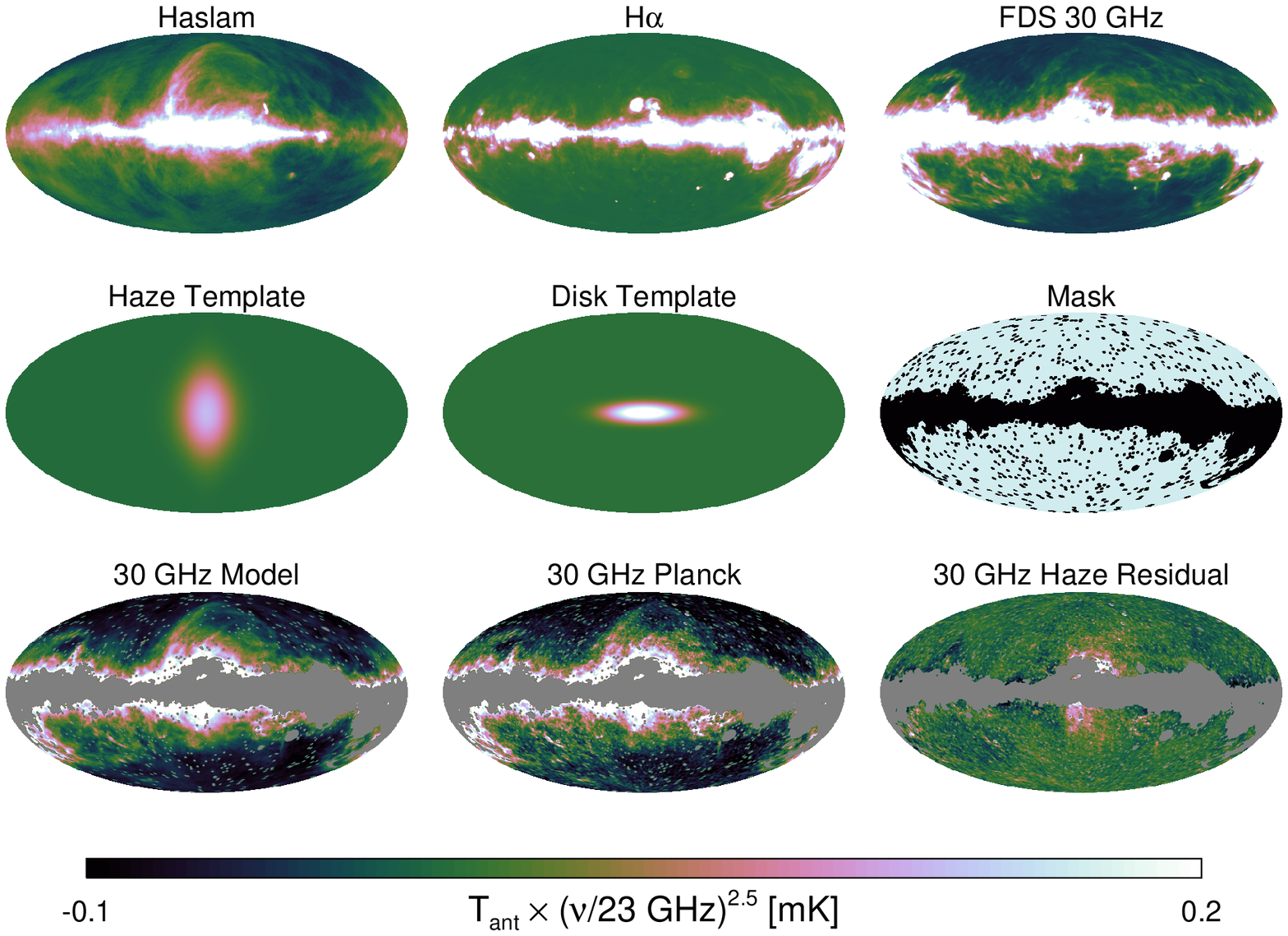}
  }
\caption{
  The templates and full-sky template fitting model
  (see \refsec{planckresults}).  
\emph{Top left}: the \citet{haslam82} 408\,MHz map. 
\emph{Top middle}: the \citet{finkbeiner03} H$\alpha$ map. 
\emph{Top right}: the \citet{finkbeiner99} dust prediction at the \planck\ 30\,GHz channel. 
\emph{Middle left}: the elliptical Gaussian haze template. 
\emph{Center}: the elliptical Gaussian disk template. 
\emph{Middle right}: the mask used in the fit. 
\emph{Bottom left}: the best fit template linear combination model at \planck\ 30\,GHz. 
\emph{Bottom middle}: the CMB-subtracted \planck\ data at 30\,GHz. 
\emph{Bottom right}: the \planck\ 30\,GHz data minus the 30\,GHz model with the haze template
  component added back into the map.
}
\label{fig:templates}
\epm

\bpm
  \centerline{
    \includegraphics[width=0.49\textwidth]{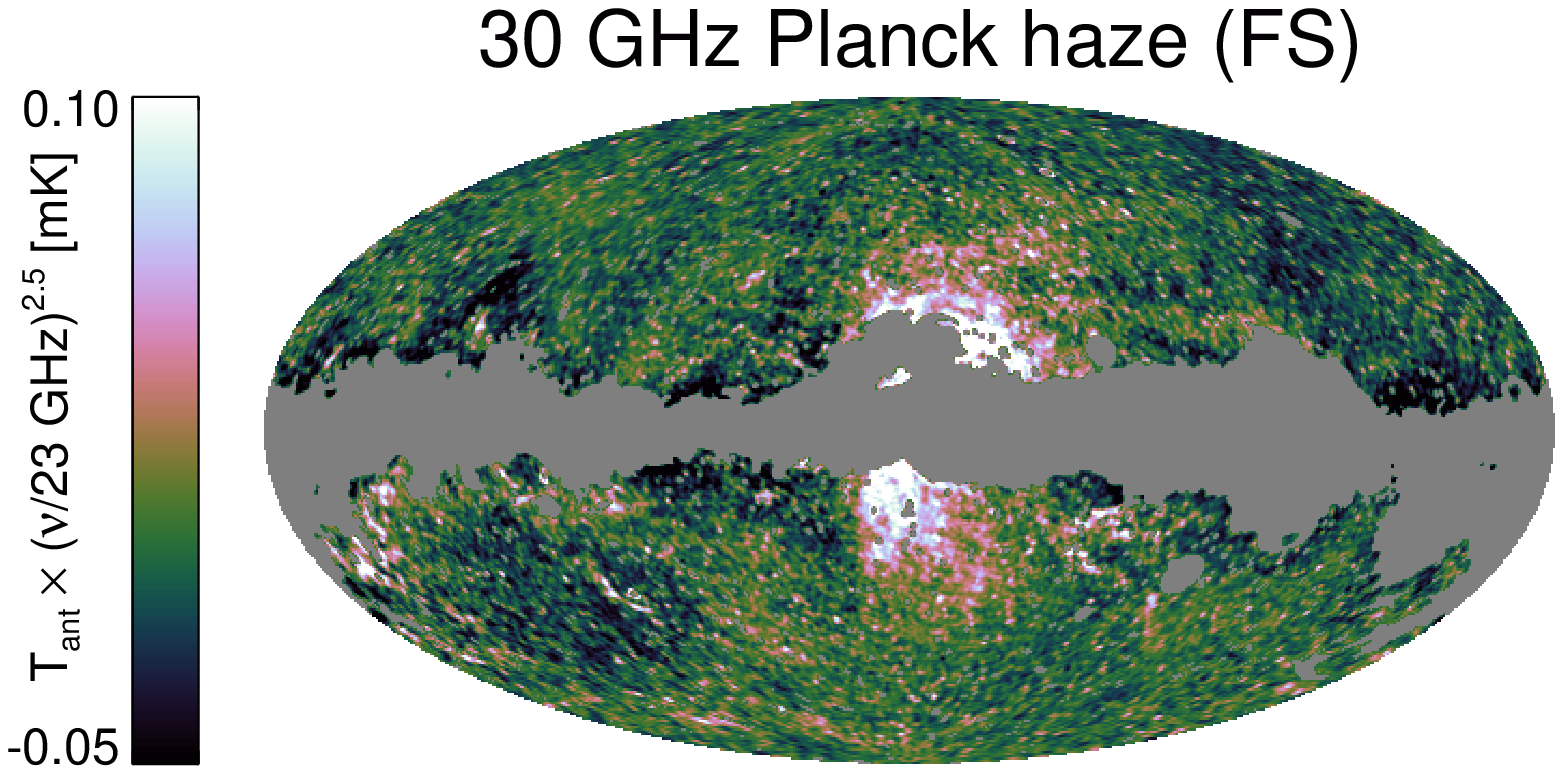}
    \includegraphics[width=0.49\textwidth]{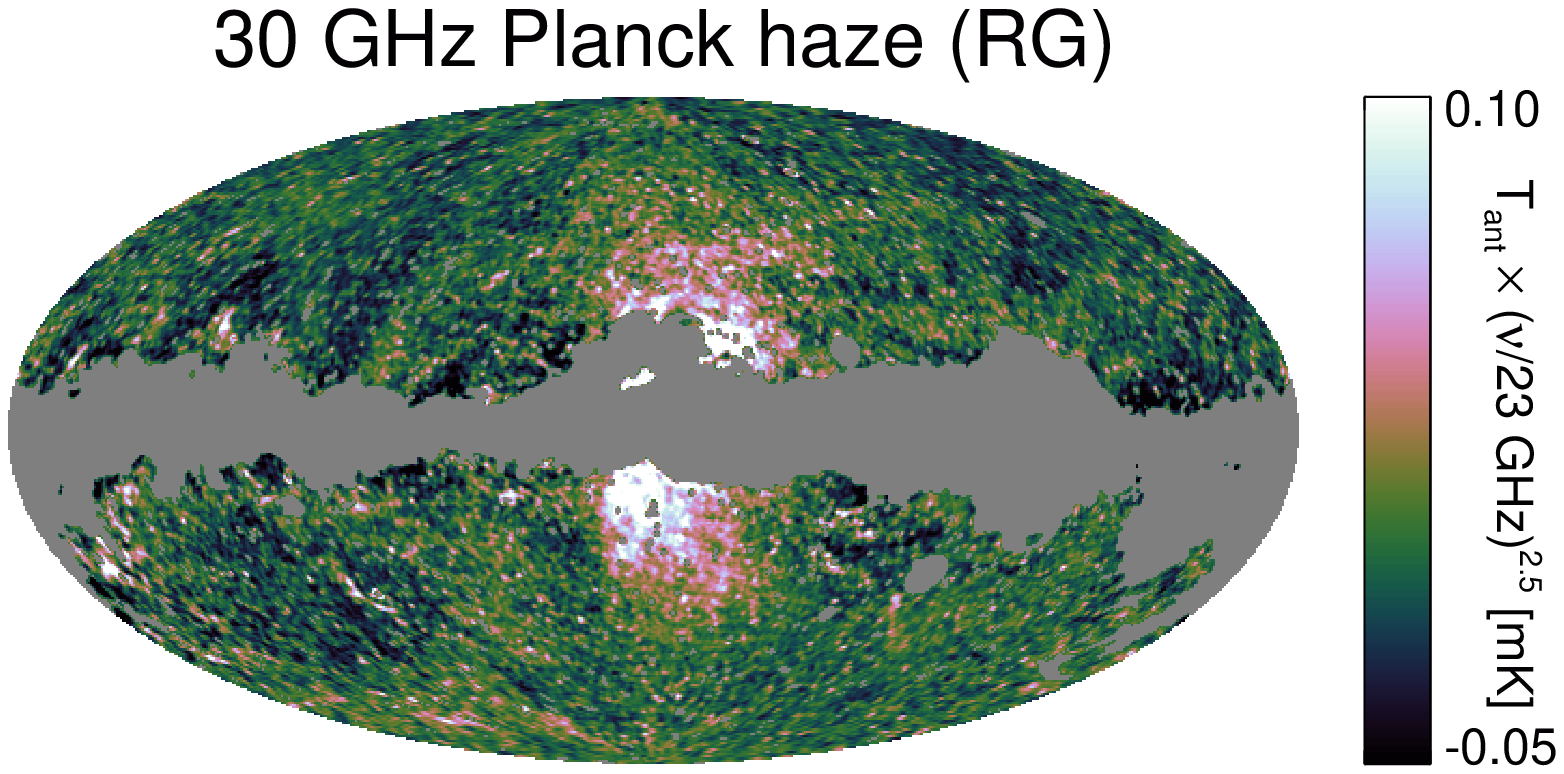}
  }
  \centerline{
    \includegraphics[width=0.49\textwidth]{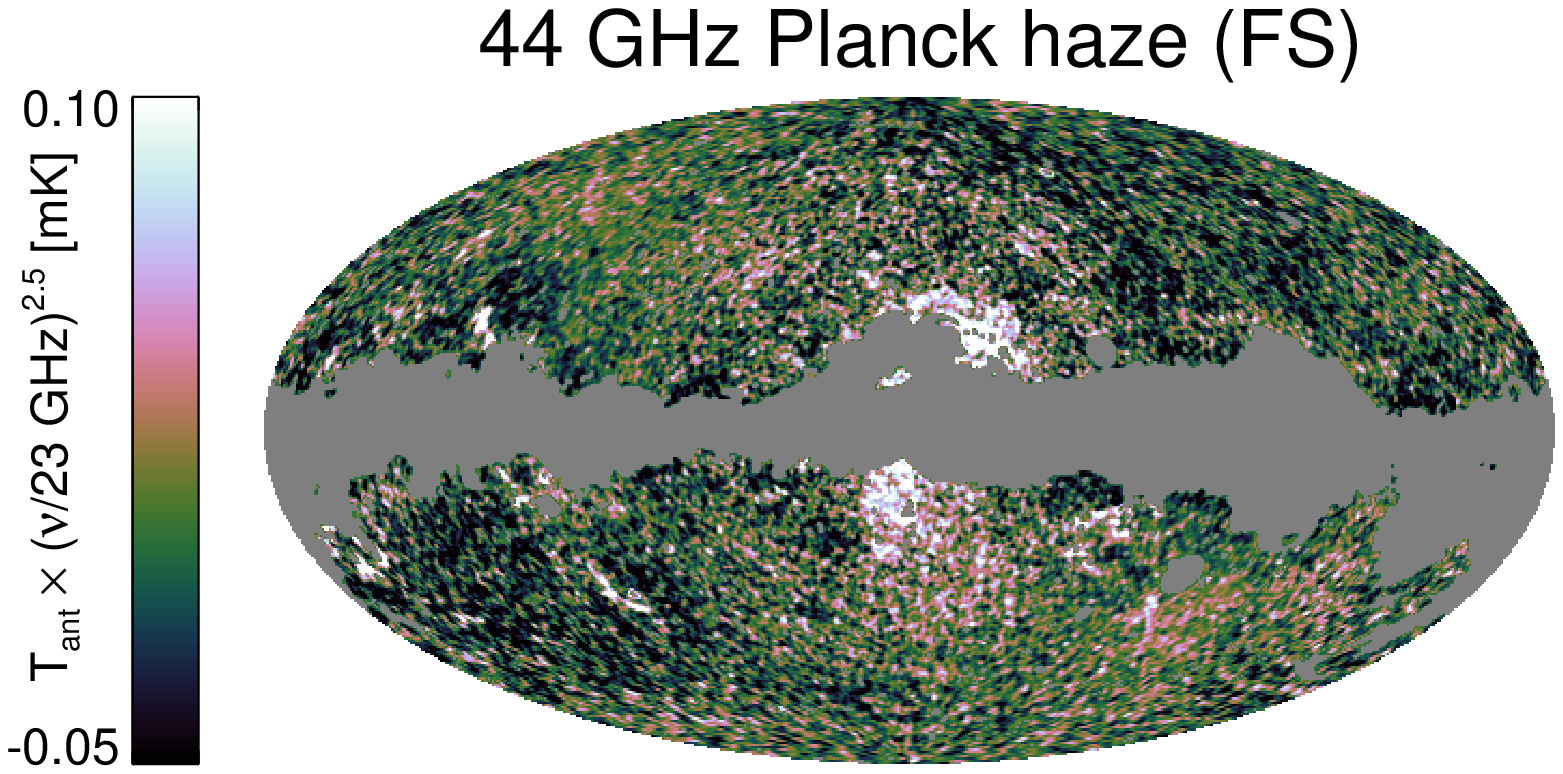}
    \includegraphics[width=0.49\textwidth]{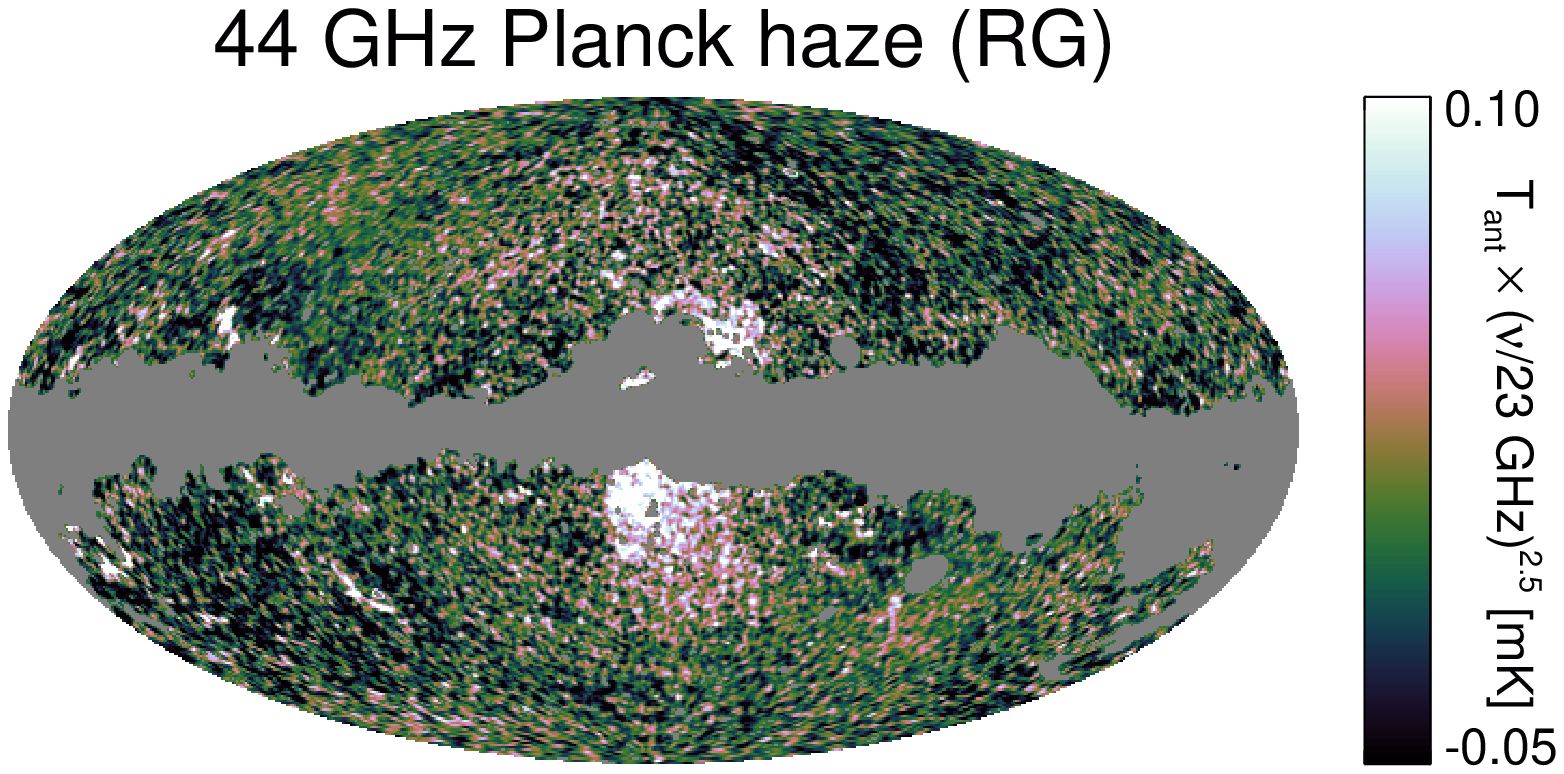}
  }
\caption{
  \emph{Left column}: the \planck\ haze (i.e., the same as the bottom
  right panel of \reffig{templates}), for the \planck\ 30 and 44\,GHz
  channels using a full-sky template fit to the data.  A scaling of
  $\nu^{2.5}$ yields roughly equal brightness residuals indicating
  that the haze spectrum is roughly $T_{\nu} \propto \nu^{-2.5}$,
  implying that the electron spectrum is a very hard $dN/dE_e \propto
  E^{-2}$.  Note that  the haze appears more elongated in latitude than
  longitude by a factor of two, which is roughly consistent
  with the \textit{Fermi} gamma-ray haze/bubbles \citep{dobler10}.
  For frequencies above $\sim$\,40\,GHz, striping in the HFI channels
  (which contaminates our CMB estimate) begins to dominate over the
  haze emission.  \emph{Right column}: the same but for the
  ``regional'' fits described in \refsec{planckresults}.  The overall
  morphology of the haze is the same, but the residuals near the mask
  and in the Ophiucus complex in the north GC are improved.  }
\label{fig:fs-haze}
\epm

Figure~\ref{fig:templates} presents the templates and mask
used for the \planck\ analysis, together with the 
CMB-subtracted data and best fit template model
at 30\,GHz.   We also show the full-sky
(i.e., unrestricted in $l$ and $b$) haze residual, defined as
\be
  \mathcal{R}^{\rm H}_{\nu} = \vec{d}_{\nu} - \vec{a}_{\nu}\cdot{\bf P}
  + {a}^{\rm H}_{\nu} \cdot  \vec{h}, 
\ee
where $\vec{h}$ is the haze template defined in \refsec{data}.  The
haze is {clearly} present in the \planck\ data set and,
 as illustrated in \reffig{fs-haze} ({left column}),
scaling each residual by $\nu^{2.5}$ yields roughly equal brightness per
frequency band
indicating that the spectrum is approximately  $T^{\rm H}_{\nu} \propto
\nu^{-2.5}$. A more detailed measurement of the spectrum will be given
in \refsec{specandmorph}.  It is also interesting to note that the
morphology does not change significantly with frequency (although
striping in the \textit{Planck} HFI maps used to form the CMB
estimate is a significant contaminant at frequencies above $\sim$\,40\,GHz) 
indicating that the spectrum of the haze emission is roughly
constant with position.

The haze residual is most clearly visible in the southern GC region, but
we note that our assumption of uniform spectra across the sky does
leave some residuals around the edge of the mask and in a few
particularly bright free-free regions.  However, while our imperfect
templates and assumptions about uniform spectra have done a remarkable
job of isolating the haze emission (96\% of the total variance is
removed in the fit at \textit{Planck} 30\,GHz), we can more effectively
isolate the haze by subdividing the sky into smaller regions as
described in \refsec{tempfit}.
The resultant full-sky haze residual is shown in \reffig{fs-haze}.
With this fit, the residuals near the mask are cleaner and we have
done a better job in fitting the difficult Ophiucus region in the northern
GC, though striping 
again becomes a major contaminant for frequencies
above $\sim$\,40\,GHz.

\subsubsection{\commander}
\label{sec:planckgibbs}

\bpm
  \centerline{
    \includegraphics[width=0.49\textwidth]{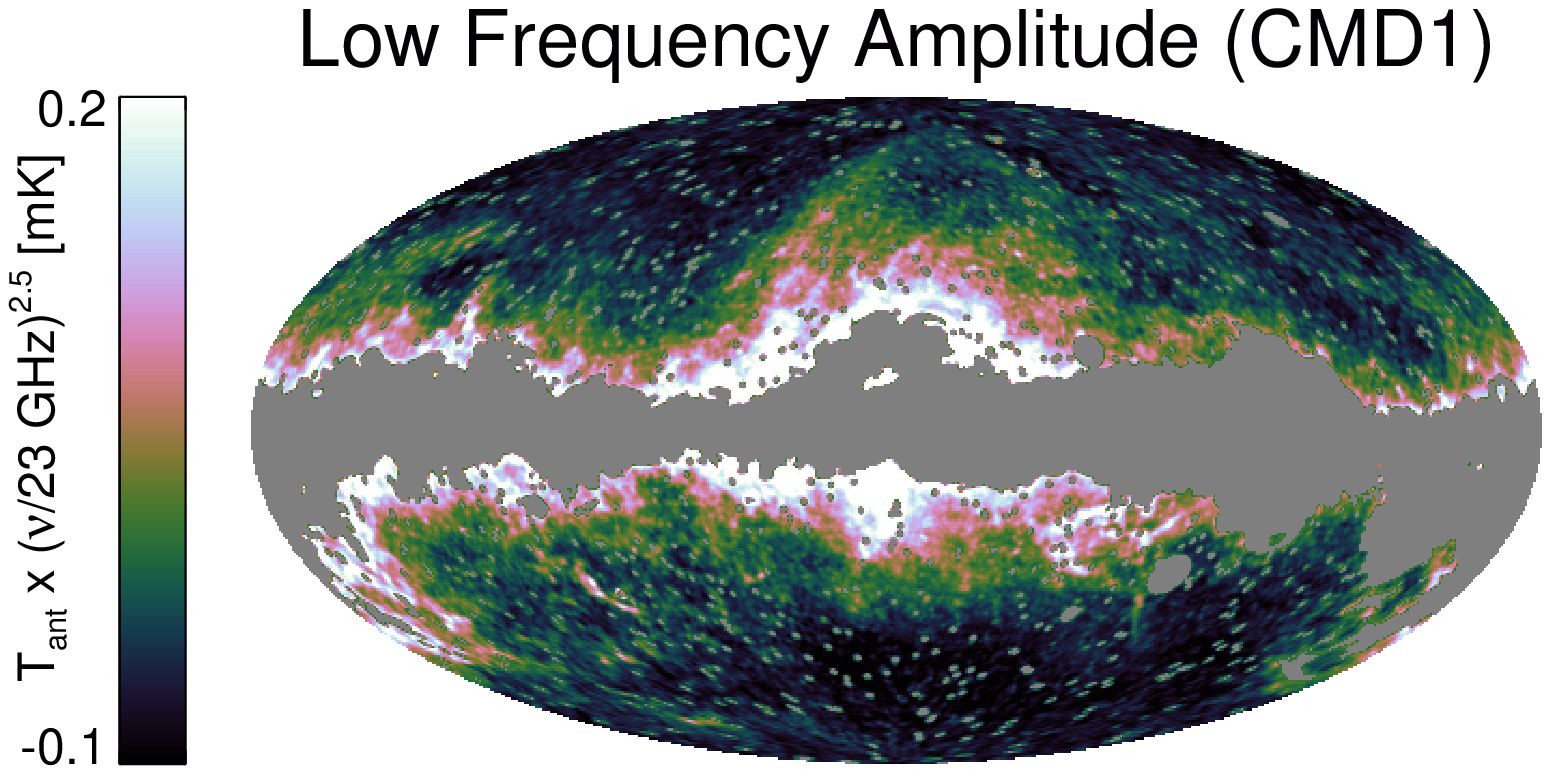}
    \includegraphics[width=0.49\textwidth]{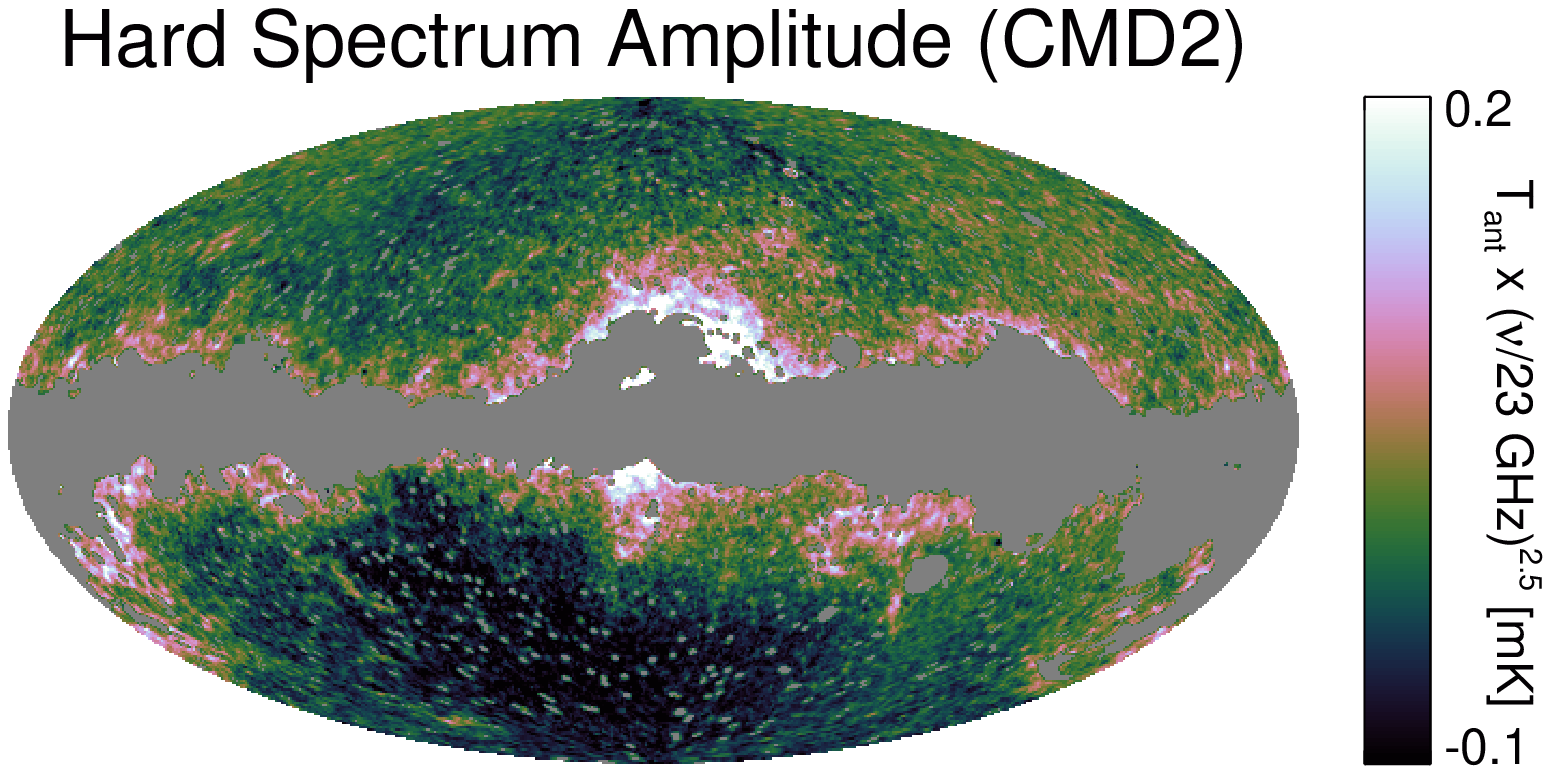}
  }
  \centerline{
    \includegraphics[width=0.49\textwidth]{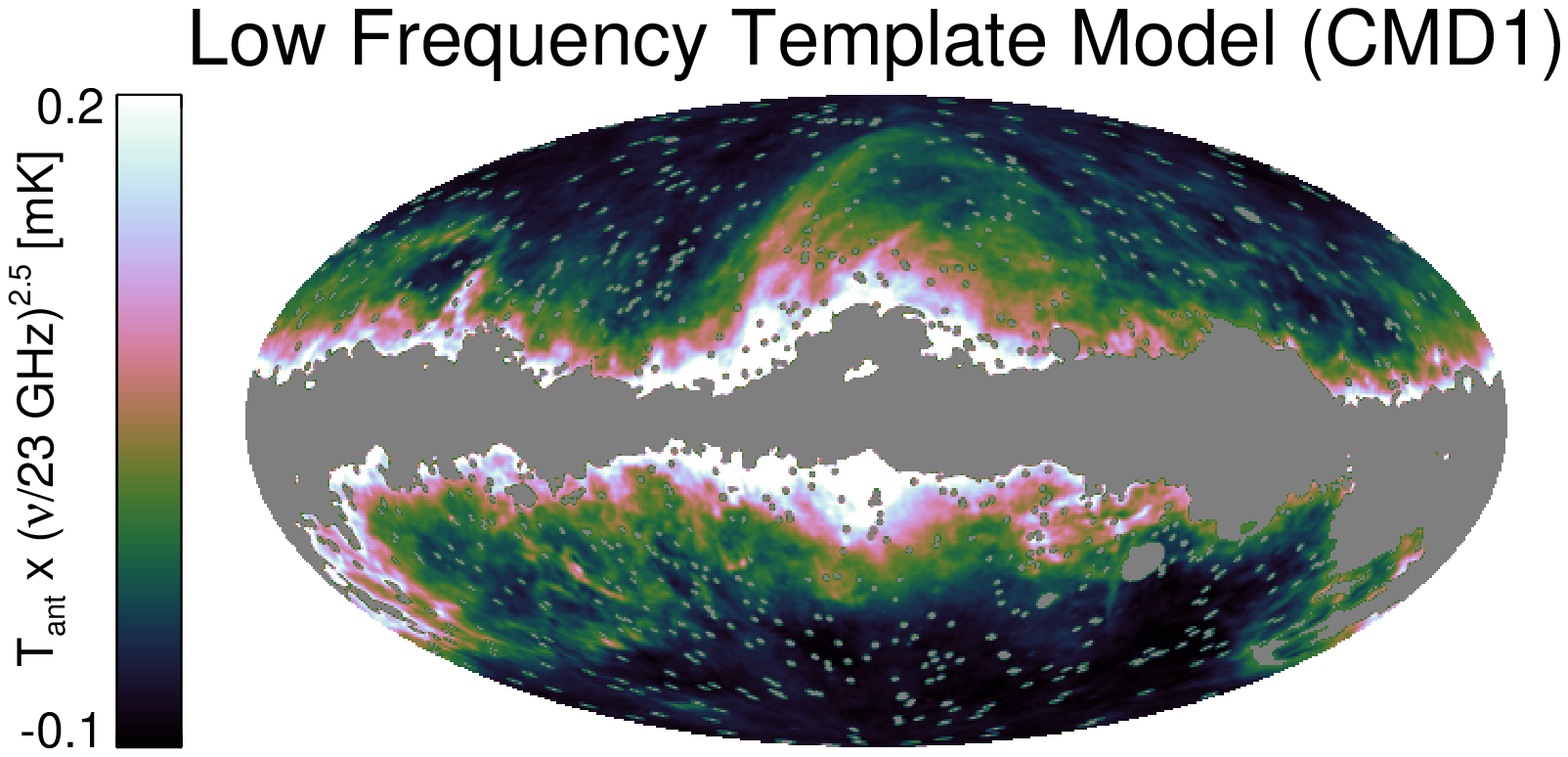}
    \includegraphics[width=0.49\textwidth]{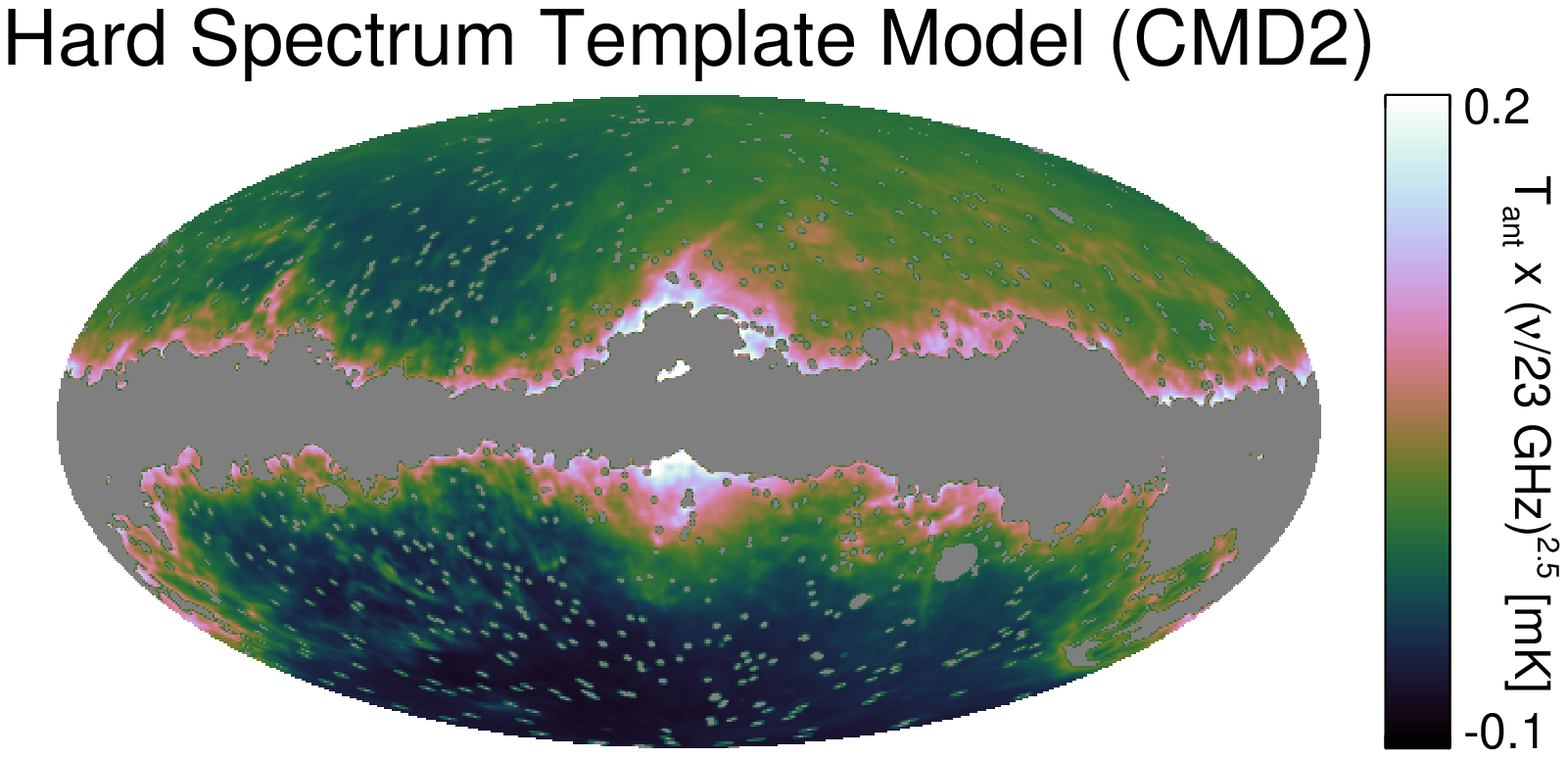}
}
  \centerline{
    \includegraphics[width=0.49\textwidth]{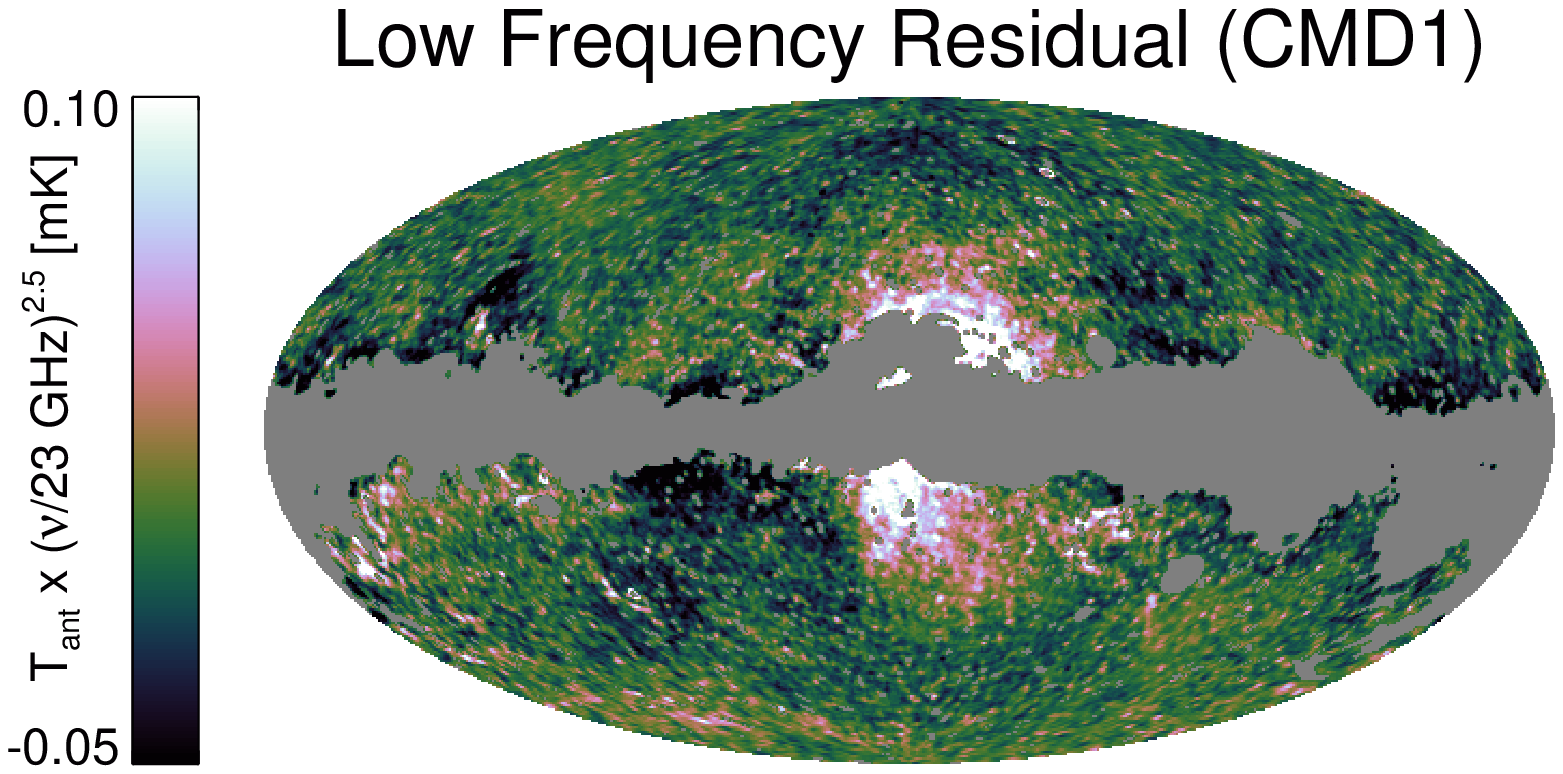}
    \includegraphics[width=0.49\textwidth]{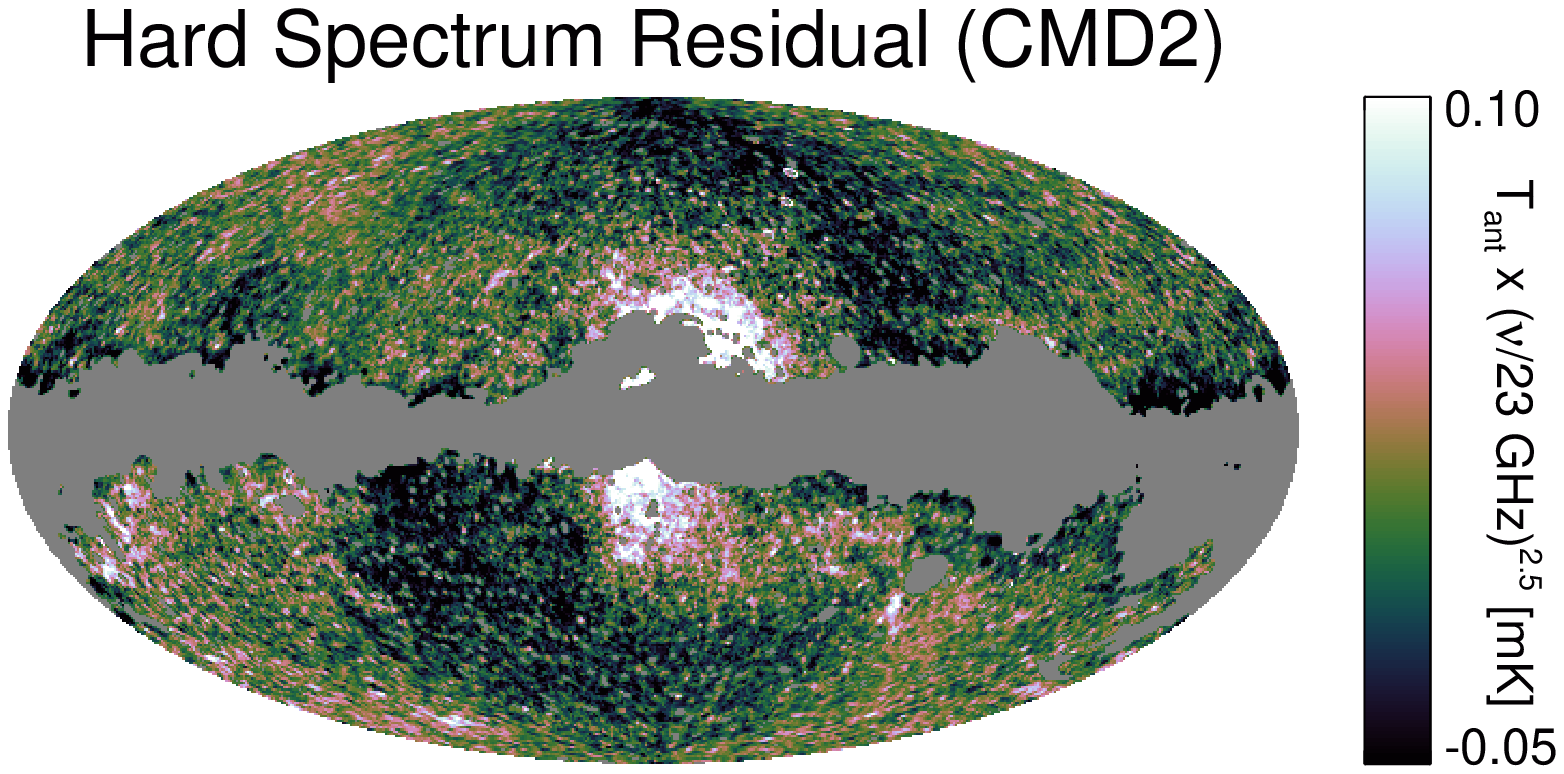}
  }
\caption{
\emph{Left column, top}: The recovered amplitude of
  the low-frequency component at 23\,GHz from our simplest
  \commander\ fit to the \planck\ data alone, CMD1.  
  As shown in \citet{pietrobon11}, while this model provides an
  excellent description of the data, this low-frequency component is
  actually a combination of free-free, spinning dust, and synchrotron
  emission (\emph{top}).  
\emph{Left column, middle}: a four-component
  template model of this component (see \reftab{planck_regression}).
\emph{Left column, bottom}: The haze residual.  The residuals are small
  outside the haze region indicating that the templates are a
  reasonable morphological representation of the different components
  contained in the \commander\ solution.  
  The haze residual is strikingly similar to that found for the
  template-only approach in \reffig{fs-haze} (though there does seem
  to be a residual dipole in the \commander\ solution).  
\emph{Right column}: The same, but for the CMD2 low-frequency, hard spectrum
  component.  While there is still some leakage of dust-correlated
  emission in the solution, the softer synchrotron emission (mostly
  correlated with the 408\,MHz template [see \reffig{softsynch_cmd}]) has been
  separated by \commander.  The resultant map is dominated by
  free-free and the haze emission and the regressed haze residual
  (\emph{bottom panel}) shows morphology very similar to both the
  template fitting and CMD1 results indicating that the haze has been
  effectively isolated.  }
\label{fig:planck_regression}
\epm

Figure \ref{fig:planck_regression} presents the results of our CMD1
\commander\ fit and the subsequent post-processing.  As noted
previously, this very simple model provides an adequate description of
the data with a mean $\chi^2$ of 18.4 (7 d.o.f.) outside the mask, despite 
the fact that the low-frequency component
is really an aggregate of several different emission mechanisms, as
shown by \citet{pietrobon11}.  It is visually apparent that the
low-frequency amplitude is highly correlated with thermal dust
emission in some regions, suggesting a dust origin for some of this
emission (e.g., spinning dust).  Finally, features that are well known
from low-frequency radio surveys, such as Loop~I, are also visible,
implying a synchrotron origin, with a spectral index closer to
$\beta_{\rm S}=-3$.  The coefficients of the post-processing
template-based fit described in \refsec{gibbssamp} are given in
\reftab{planck_regression} and show a strong positive correlation with
each template.

\begin{table*}[t]
\caption{
  Regression coefficients of the \commander\ foreground amplitude
  maps.}
\label{tab:planck_regression}
\begin{tabular}{clcccc}
        \noalign{\hrule\vskip2pt\hrule\vskip 4pt}
 & & \multicolumn{4}{c}{\bf Fit coefficient}\\
\noalign{\vskip-3pt}
  {\bf Fit type}& \multicolumn{1}{c}{\bf Data sets}& \multicolumn{4}{c}{\hrulefill}\\ 
  & & H$\alpha$ [mK/R] & FDS [mK/mK] & Haslam [mK/K] & Haze [mK/arbitrary] \\
  \hline\\
  CMD1 & \planck\ 30--353\,GHz &
$2.8\times10^{-3}\pm2.0\times10^{-4}$ &
$1.9\pm4.3\times10^{-2}$ &
$1.6\times10^{-6}\pm4.4\times10^{-8}$ &
$6.0\times10^{-2}\pm3.4\times10^{-3}$ \\[1em]
  CMD2 & \parbox{3.0cm}{\planck\ 30--353\,GHz,\\ \textit{WMAP}, Haslam} &
$3.3\times10^{-3}\pm3.9\times10^{-4}$ &
$1.0\pm8.4\times10^{-2}$ &
$2.4\times10^{-9}\pm8.8\times10^{-8}$ &
$5.7\times10^{-2}\pm6.7\times10^{-3}$ \\[1em]
    \noalign{\vskip 3pt\hrule}
\end{tabular}
\end{table*}

As with the template fitting case, we see from
\reffig{planck_regression} that the post-processing residuals for the
low-frequency CMD1 component are low except towards the Galactic
centre where the haze is clearly present, implying that it is emission
with a distinct morphology compared to the dust, free-free, and soft
synchrotron emission.  Furthermore, the morphology is strikingly
similar to the template fitting indicating strong consistency between
the results.  Since an analogous regression cannot be performed on the
spectral-index map, a more flexible foreground model must be
implemented to isolate the haze spectrum.  However, the additional
model parameters require the use of external data sets.

\subsection{Results from \planck\ plus external data sets}
\label{sec:planckwmapresults}
\subsubsection{Template fitting}
\label{sec:planckwmaptemp}

\bpm
  \centerline{
    \includegraphics[width=0.49\textwidth]{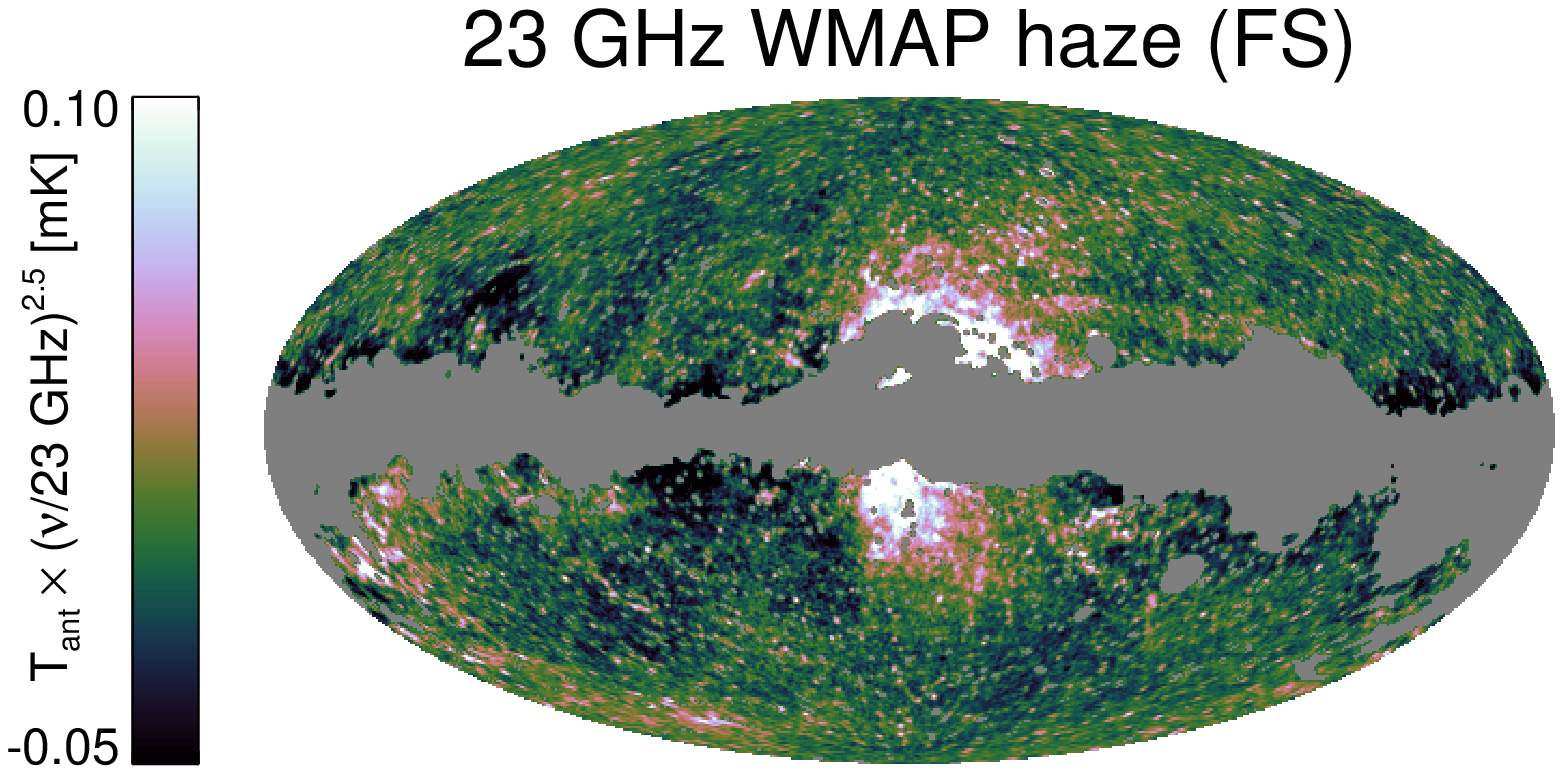}
    \includegraphics[width=0.49\textwidth]{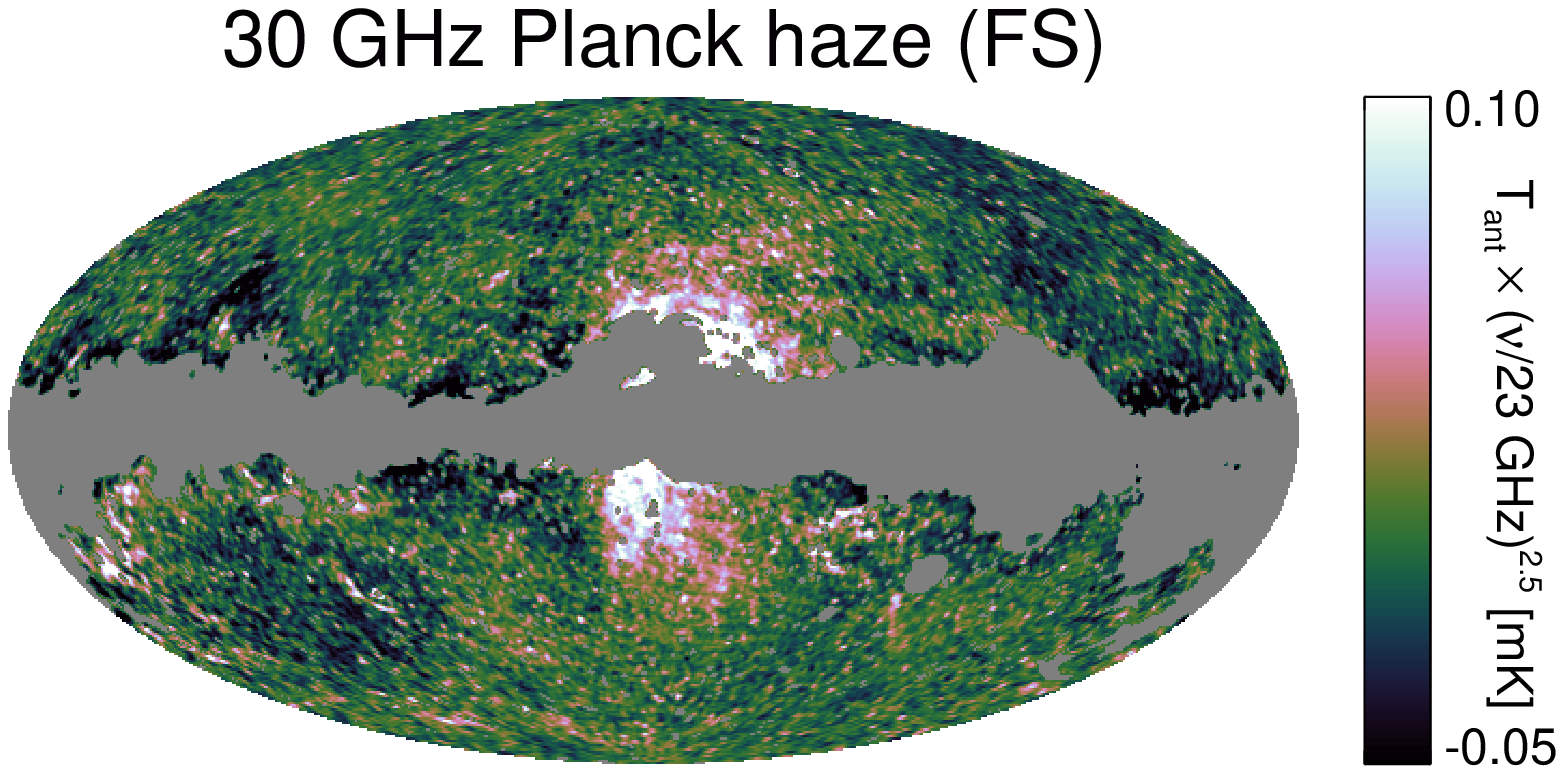}
  }
  \centerline{
    \includegraphics[width=0.49\textwidth]{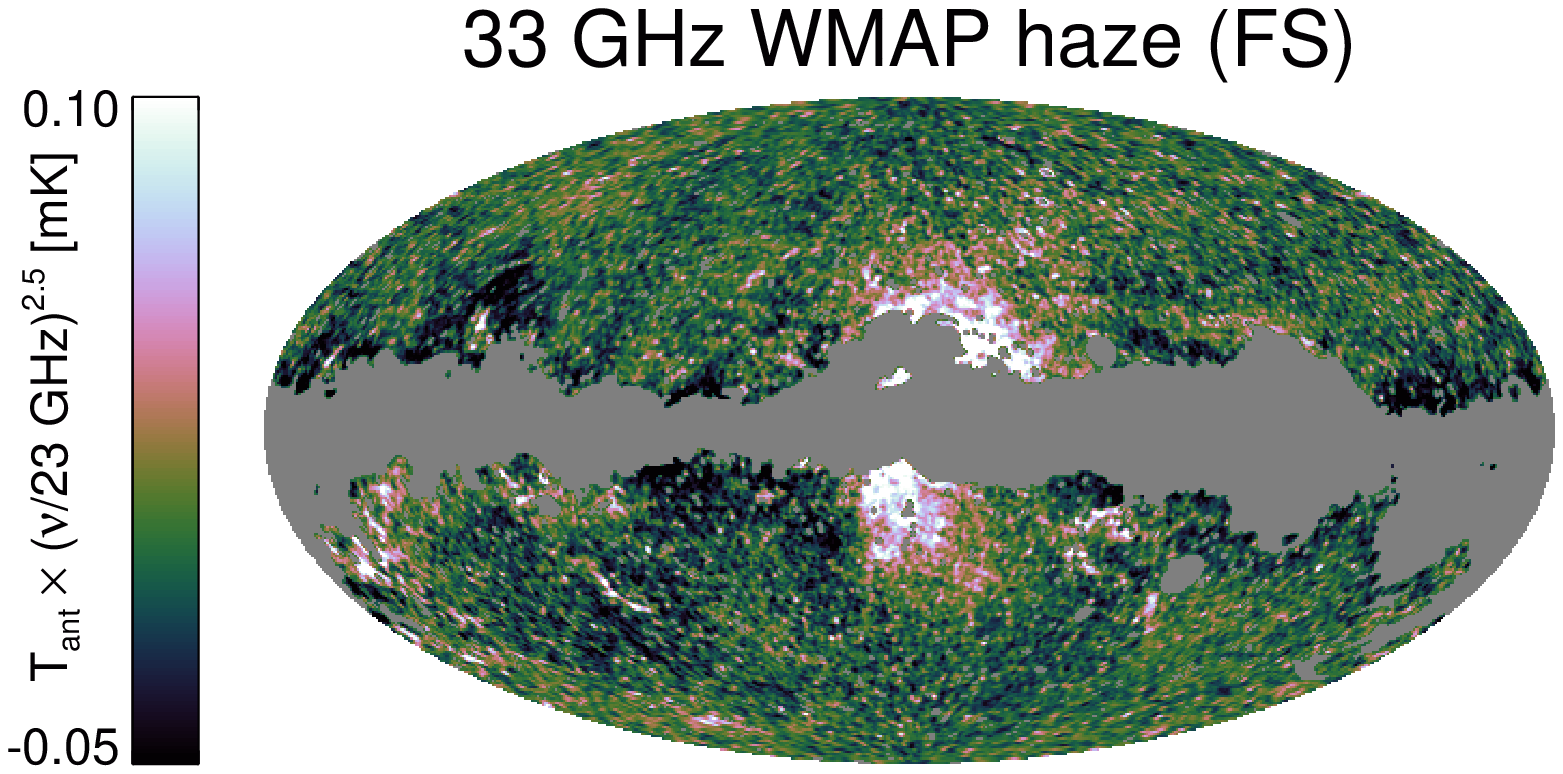}
    \includegraphics[width=0.49\textwidth]{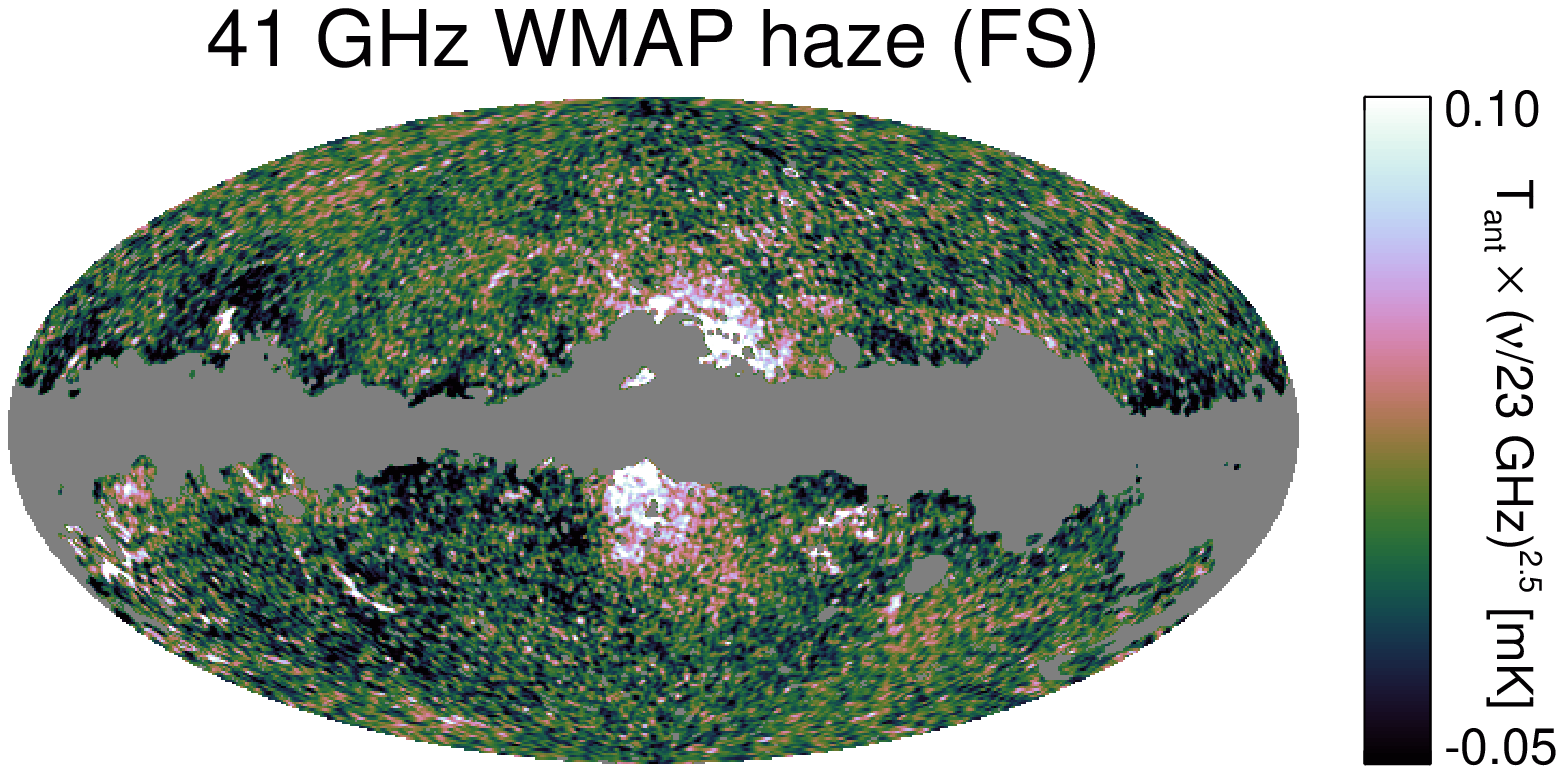}
  }
  \centerline{
    \includegraphics[width=0.49\textwidth]{haze_044_left_FS.eps}
    \includegraphics[width=0.49\textwidth]{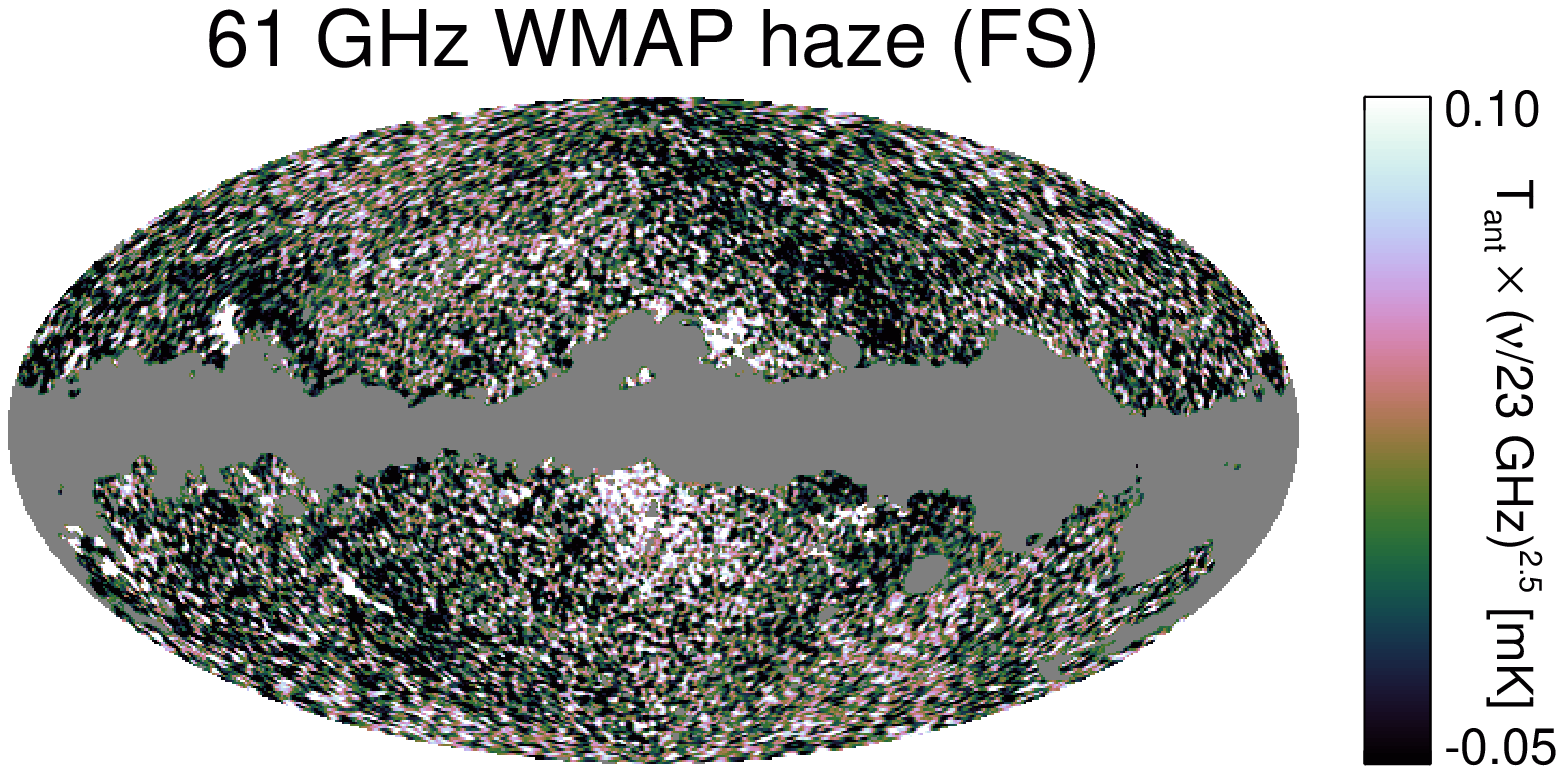}
  }
\caption{
  The microwave haze at both \textit{WMAP} and \planck\ wavelengths using a
  full-sky template fit to the data.  The morphology of the haze is
  remarkably consistent from band to band and between data sets
  implying that the spectrum of the haze does not vary significantly
  with position.  Furthermore, the $\nu^{2.5}$ scaling again yields
  roughly equal-brightness residuals indicating that the haze spectrum
  is roughly $T_{\nu} \propto \nu^{-2.5}$ through both the \planck\
  and \textit{WMAP} channels.  In addition, while striping is minimally
  important at low frequencies, above $\sim$\,40\,GHz it becomes
  comparable to, or brighter than, the haze emission (see text).  }
\label{fig:fs-haze-all}
\epm

\bpm
  \centerline{
    \includegraphics[width=0.49\textwidth]{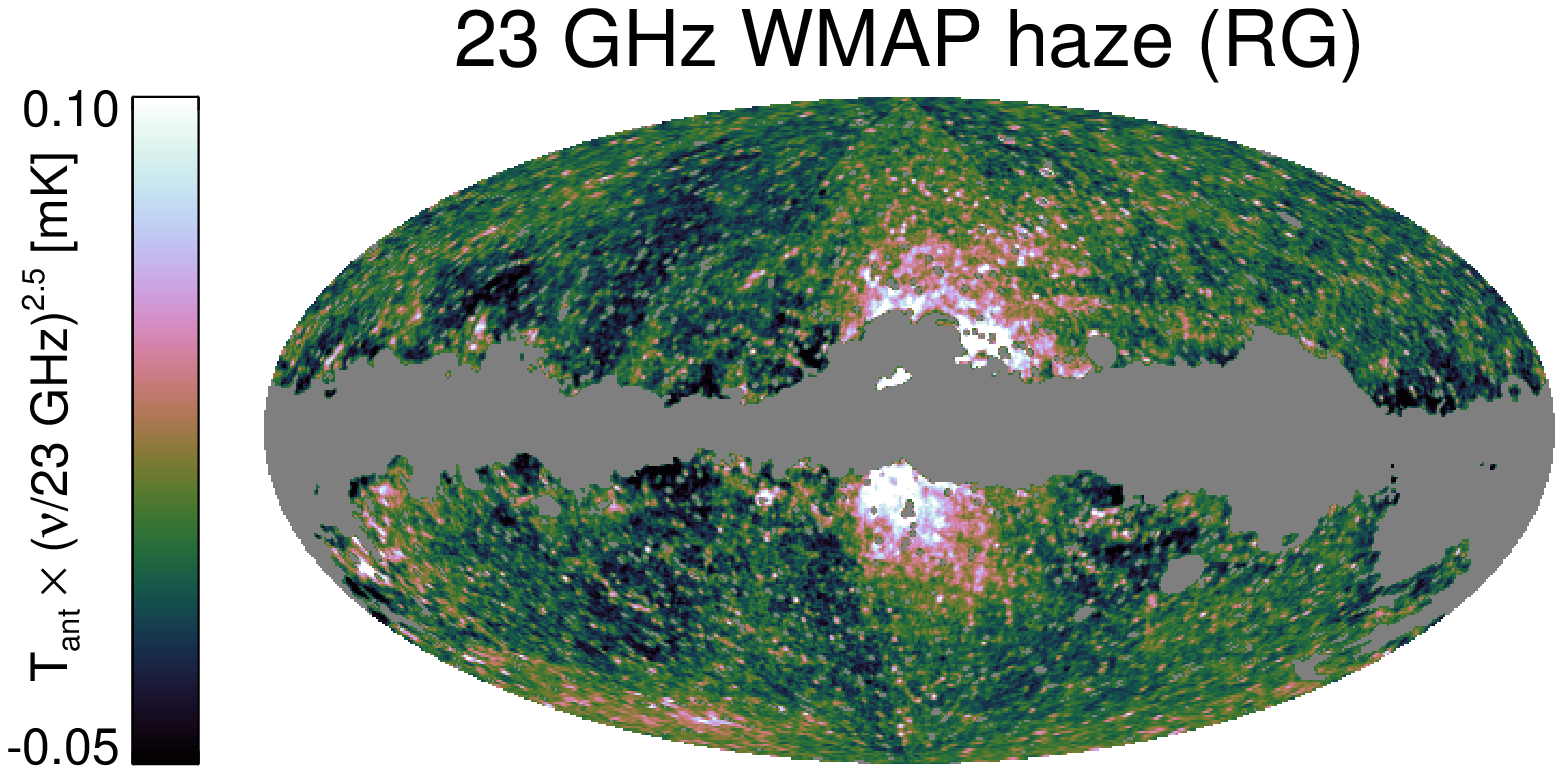}
    \includegraphics[width=0.49\textwidth]{haze_030_right_RG.eps}
  }
  \centerline{
    \includegraphics[width=0.49\textwidth]{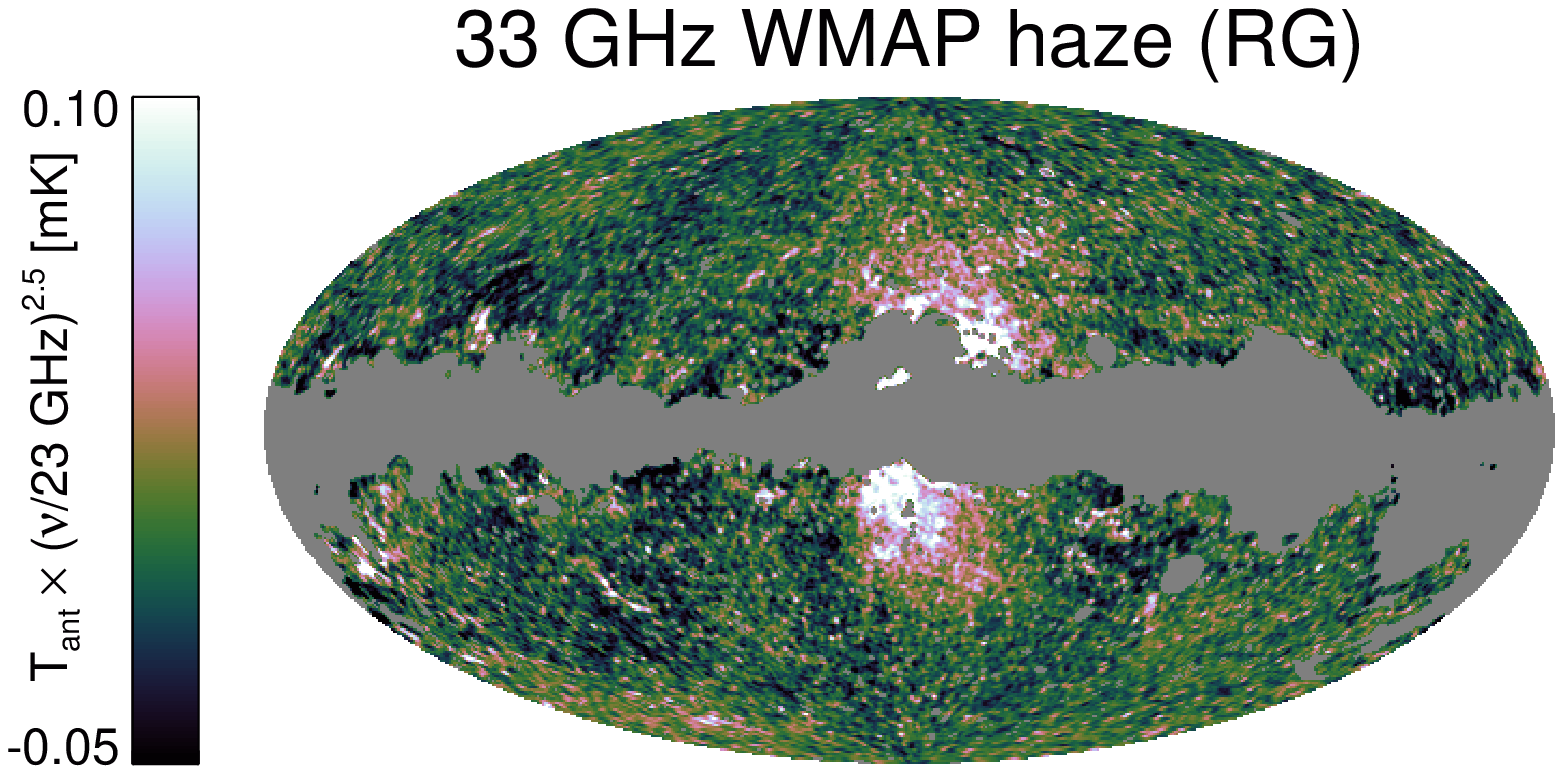}
    \includegraphics[width=0.49\textwidth]{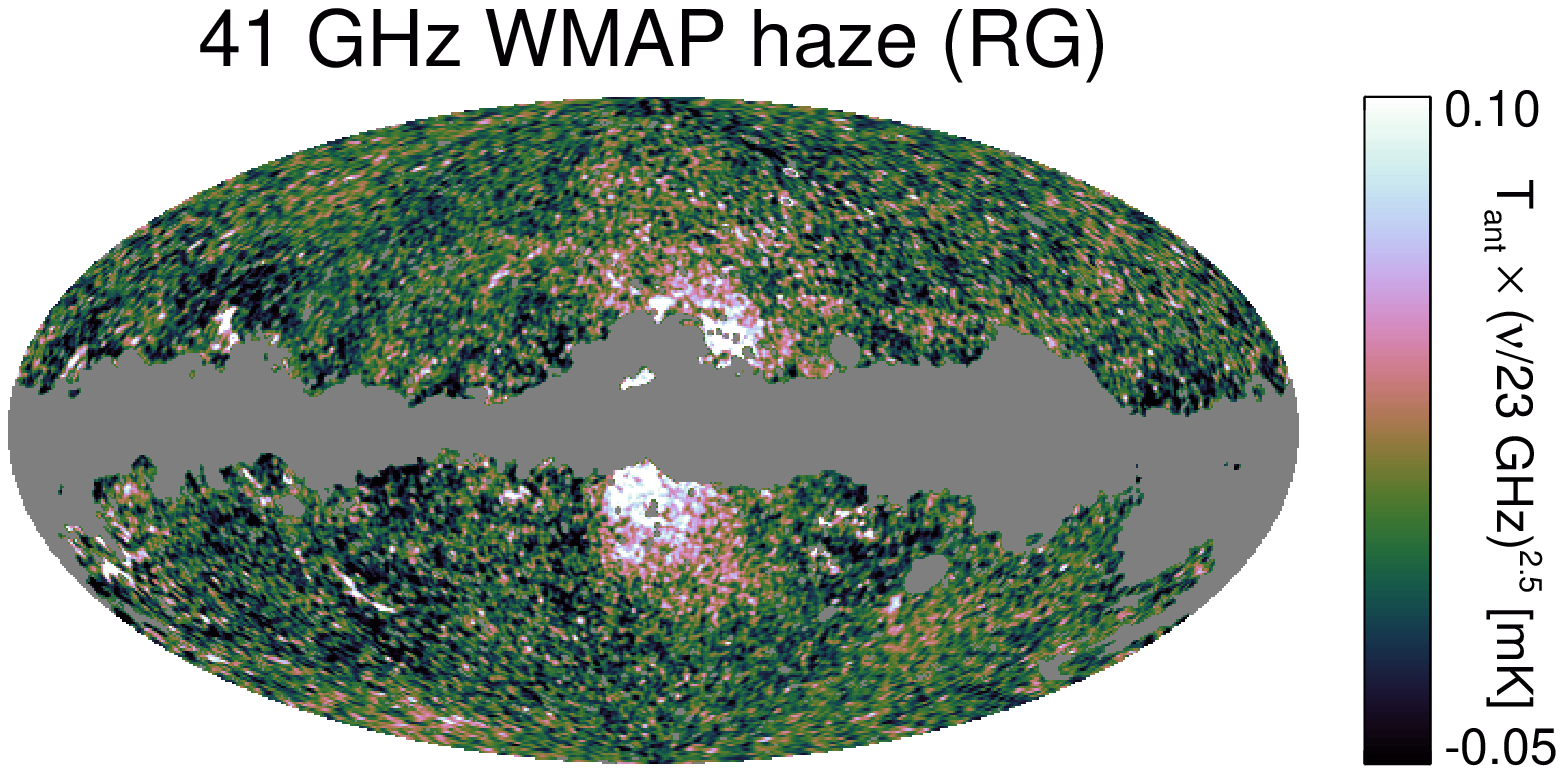}
  }
  \centerline{
    \includegraphics[width=0.49\textwidth]{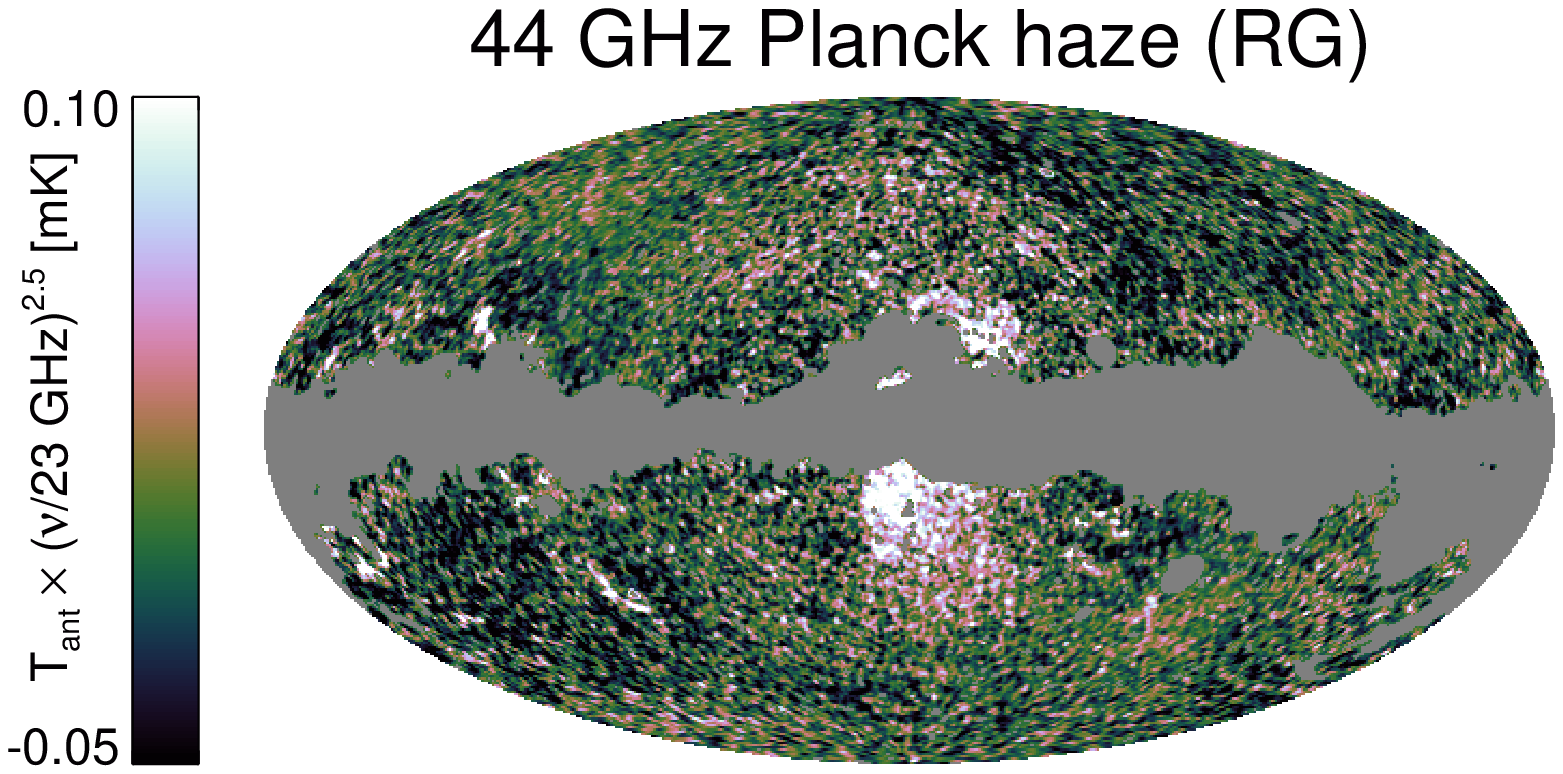}
    \includegraphics[width=0.49\textwidth]{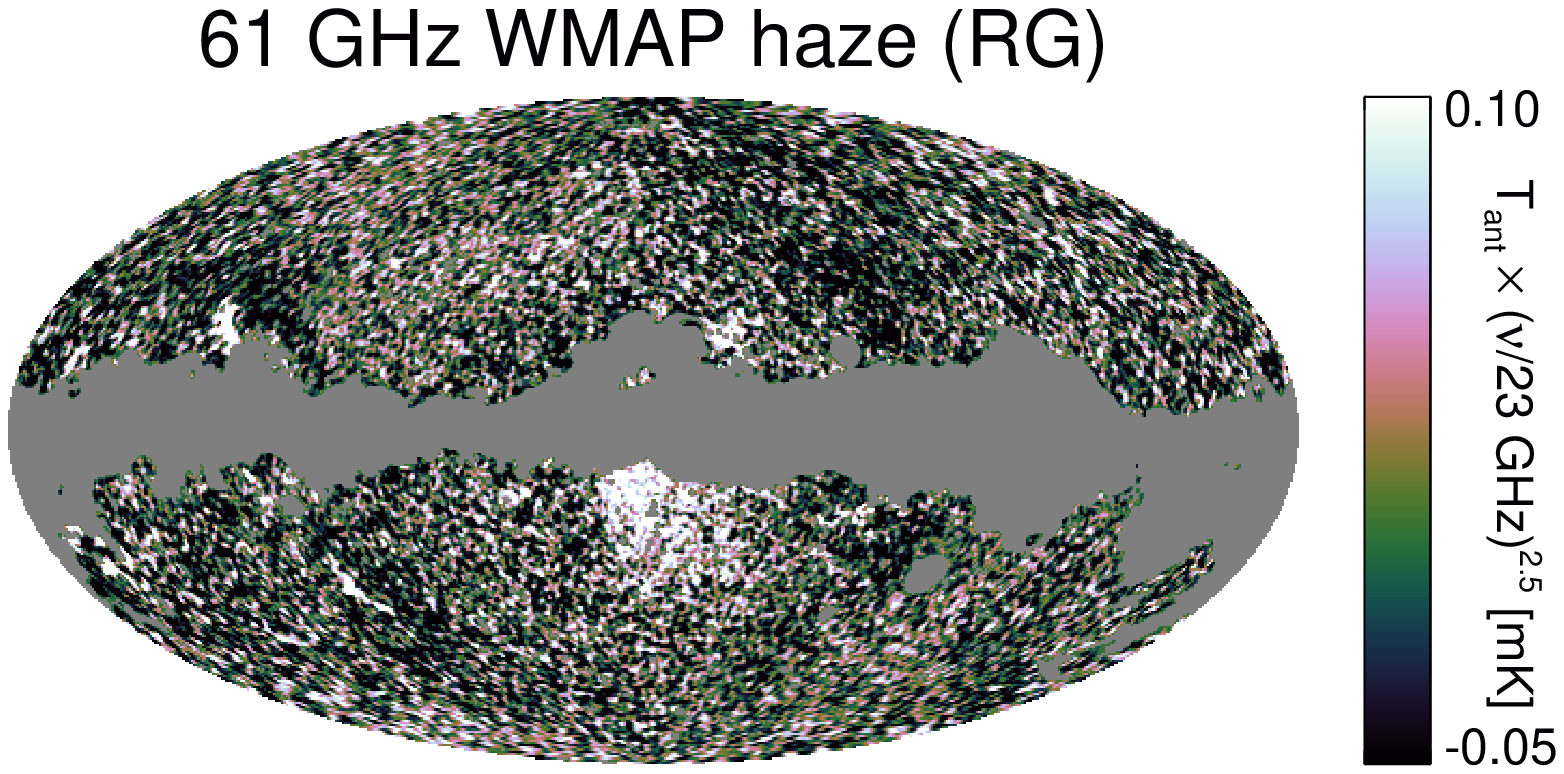}
  }
\caption{
  The same as \reffig{fs-haze-all} but using the regions defined in
  \citetalias{dobler08a}.  Clearly, the residuals near the mask are significantly reduced,
  although, as with the full-sky fits, striping in the HFI channels (which
  leaks into the CMB estimate) becomes significant above $\sim$\,40\,GHz.
  }
\label{fig:rg-haze-all}
\epm

In order to further our understanding of the spectrum and morphology
of the microwave haze component, we augment the \planck\ data with the
\textit{WMAP} 7-year data set (covering the frequency range
23--94\,GHz) and the 408\,MHz data.  For the template-fitting method,
the inclusion of the new data is trivial since Eq.~\ref{eq:tempfit}
does not assume anything about the frequency dependence of the
spectrum and each map is fit independently.  The results for the full
sky and for smaller regional fits are shown in
Figs.~\ref{fig:fs-haze-all} and \ref{fig:rg-haze-all}.  The haze
residual is present in both the \textit{WMAP} and \textit{Planck}
data, and the morphology and spectrum appear consistent between data
sets.  As before, scaling each residual by $\nu^{2.5}$ yields roughly
equal brightness per band from 23\,GHz to 61\,GHz.
Including the \textit{WMAP} data also confirms that the morphology
does not change significantly with frequency, thus implying a roughly
constant haze spectrum with position.

\subsubsection{\commander}
\label{sec:planckwmapgibbs}

\bpm
  \centerline{
    \includegraphics[width=0.49\textwidth]{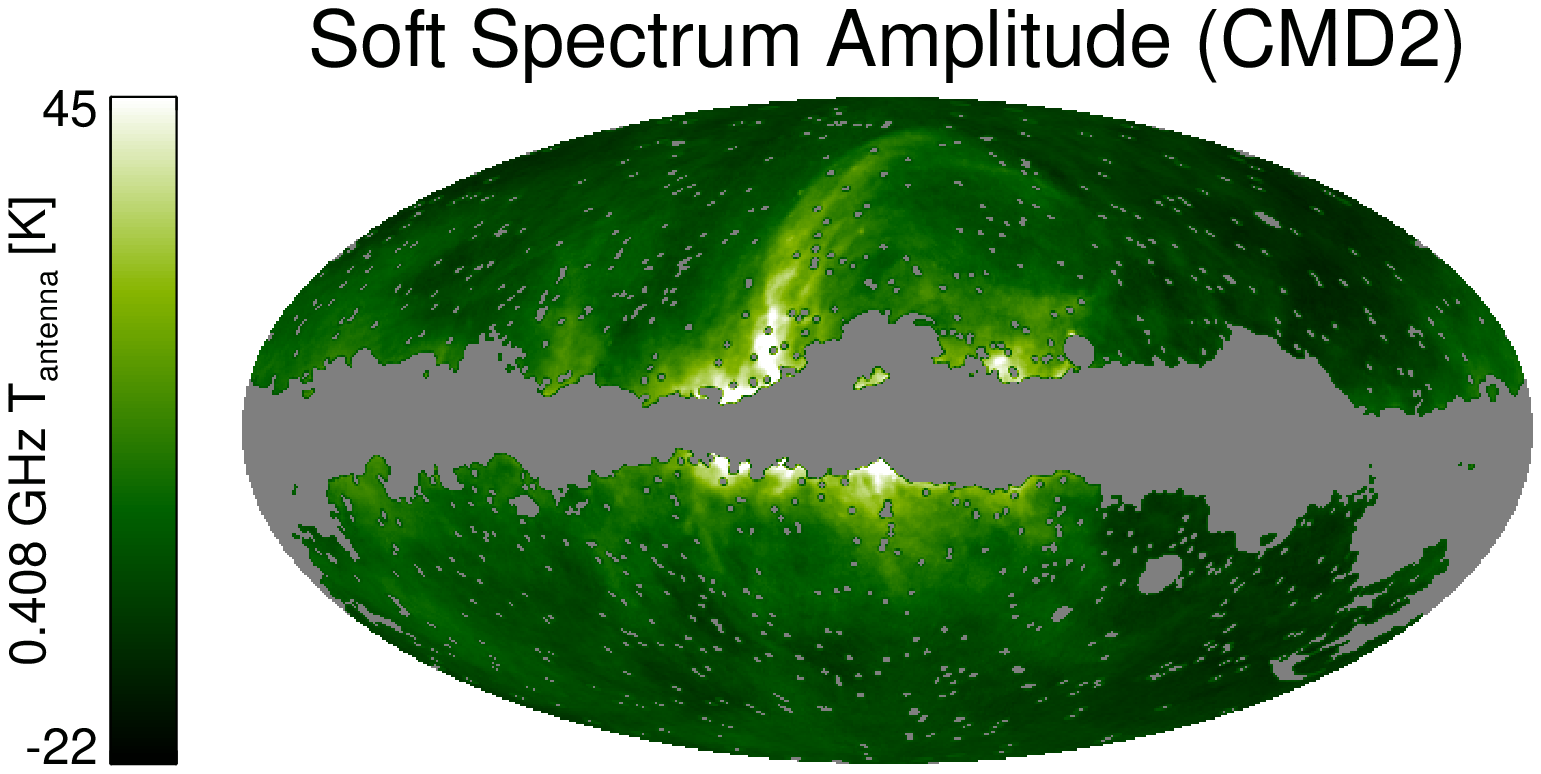}
    \includegraphics[width=0.49\textwidth]{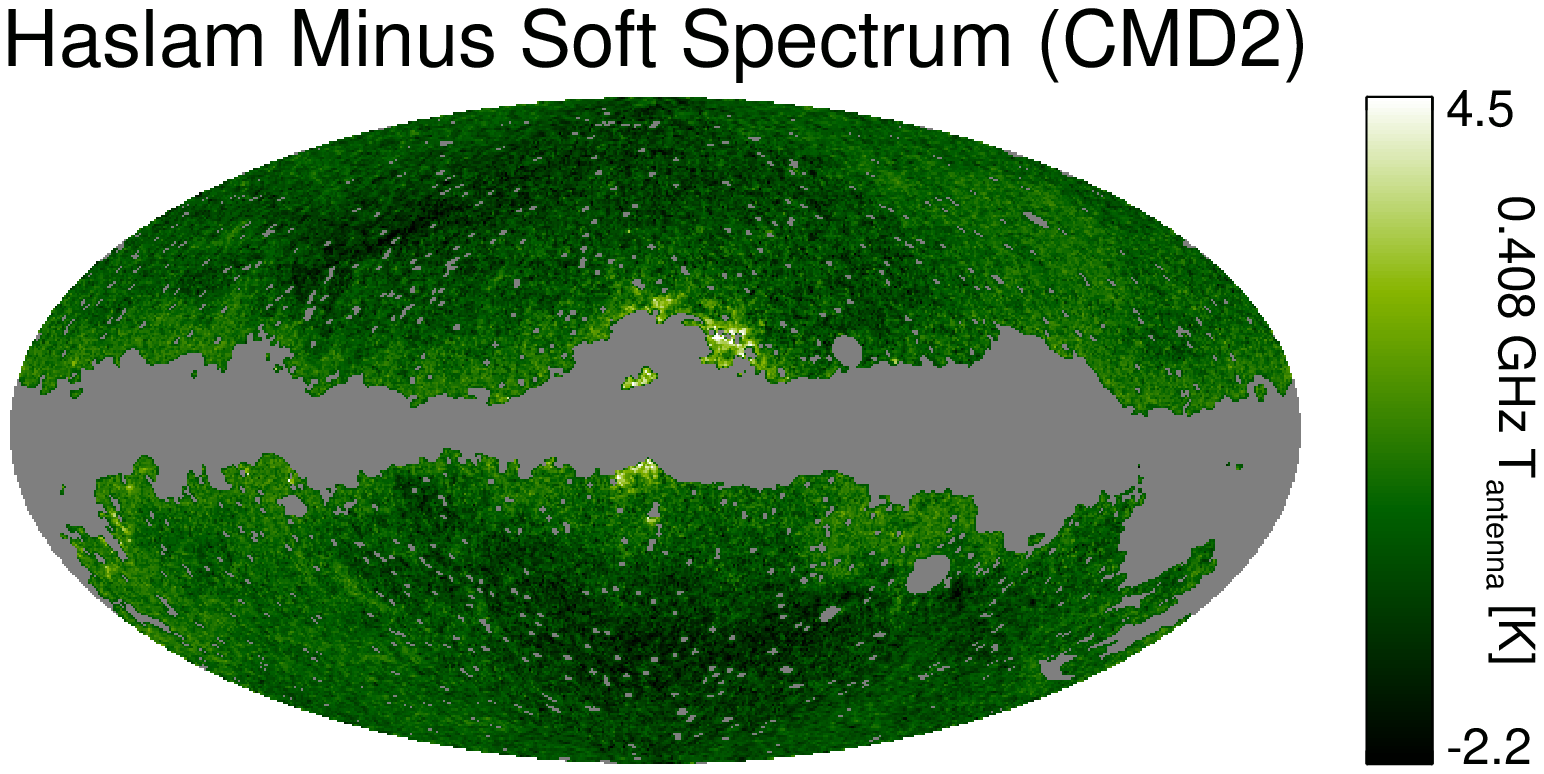}
  }
\caption{
  \emph{Left}: The soft synchrotron component at 408\,MHz from the
  \commander ~CMD2 analysis.  The map is strikingly similar to the
  Haslam map (see \reffig{templates}) indicating that soft synchrotron
  emission has a very uniform spectrum from 408\,MHz to 60\,GHz through all of
  the data sets.  
\emph{Right}: The difference between the Haslam map and
  the \commander\ solution. This is consistent with noise
  across almost the entire sky with the exception of a few bright
  free-free clouds that are present in the Haslam data at the $\sim10\%$
  level.  The lack of significant haze emission in the difference map
  (particularly in the south) is a strong indication that the haze
  region consists of both a hard \emph{and} a soft component rather than having
  a simple spatially variable spectral index.
}
\label{fig:softsynch_cmd}
\epm

Comparing the low frequency, hard spectral index \commander\ solution
at 23\,GHz obtained with this model to our previous (less flexible)
parameterisation, we find that the residuals correlated with the Haslam 408\,MHz
map are significantly reduced as shown in \reffig{planck_regression}.
\reftab{planck_regression} lists the fit coefficients in this case,
and we now find no significant correlation with the Haslam map.
As before, a template regression illustrates that the haze
residual is significant and our hard spectrum power law contains both
free-free and haze emission.\footnote{A close comparison between the
  CMD1 and CMD2 results suggests that the haze amplitude is slightly
  lower in the latter.  However, due to the flexibility of the CMD2
  model (specifically the fact that the model allows for the
  unphysical case of non-zero spinning dust in regions of negligible
  thermal dust), it is likely that some of the haze emission is being
  included in the spinning dust component.}  Furthermore,
\reffig{softsynch_cmd} illustrates that the fixed $\beta_{\rm
  S}=-3.05$ power law provides a remarkably good fit to the 408\,MHz
data. Indeed, subtracting this soft-spectrum component from the map
yields nearly zero residuals outside the mask, except for bright
free-free regions which contaminate the \citet{haslam82} map at the
$\sim10\%$ level.  It is interesting to note that this residual (as well as the
negligible Haslam-correlation coefficient in
\reftab{planck_regression}) imply that fits
assuming a constant spectral index across the sky for this correlated
emission are reasonable.
Physically, this means that electrons do
diffuse to a steady-state spectrum which is very close to $dN/dE
\propto E^{-3}$ (in agreement with the propagation models of
\citealt{strong11}).

Taken together, Figs.~\ref{fig:planck_regression}
and \ref{fig:softsynch_cmd} imply that, not only is the 408\,MHz-correlated
soft synchrotron emission consistent with a spectral index of $-3.05$
across the entire sky (outside our mask) from 408\,MHz to 60\,GHz, but the
haze region consists of both a soft and a hard component.  That
is, the haze is not a simple variation of spectral index from 408\,MHz to
$\sim20$\,GHz.  If it were, then our assumption of $\beta_{\rm S}=-3.05$
(i.e., the wrong spectral index for the haze) would yield residuals in
the difference map of \reffig{softsynch_cmd}.  The map of the harder
spectral index would ideally be a direct measurement of the haze
spectrum.  However, the signal-to-noise ratio is only sufficient to
accurately measure the spectrum in
the very bright free-free regions (e.g., the Gum Nebula).  In the fainter haze region, the
spectral index is dominated by noise in the maps.

\subsection{Spectrum and morphology}
\label{sec:specandmorph}
\bpm
  \centerline{
    \includegraphics[width=0.33\textwidth]{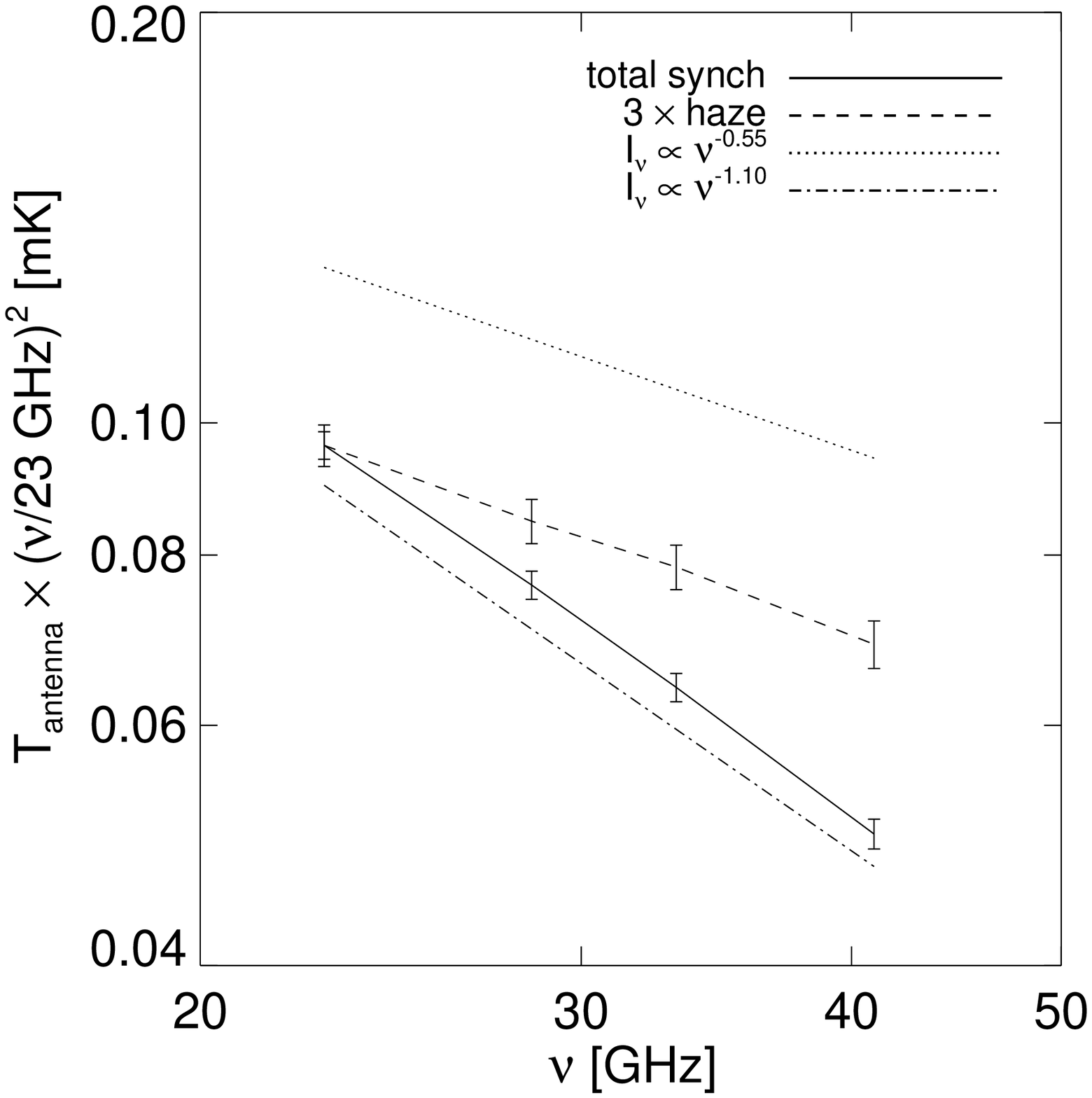}
    \includegraphics[width=0.33\textwidth]{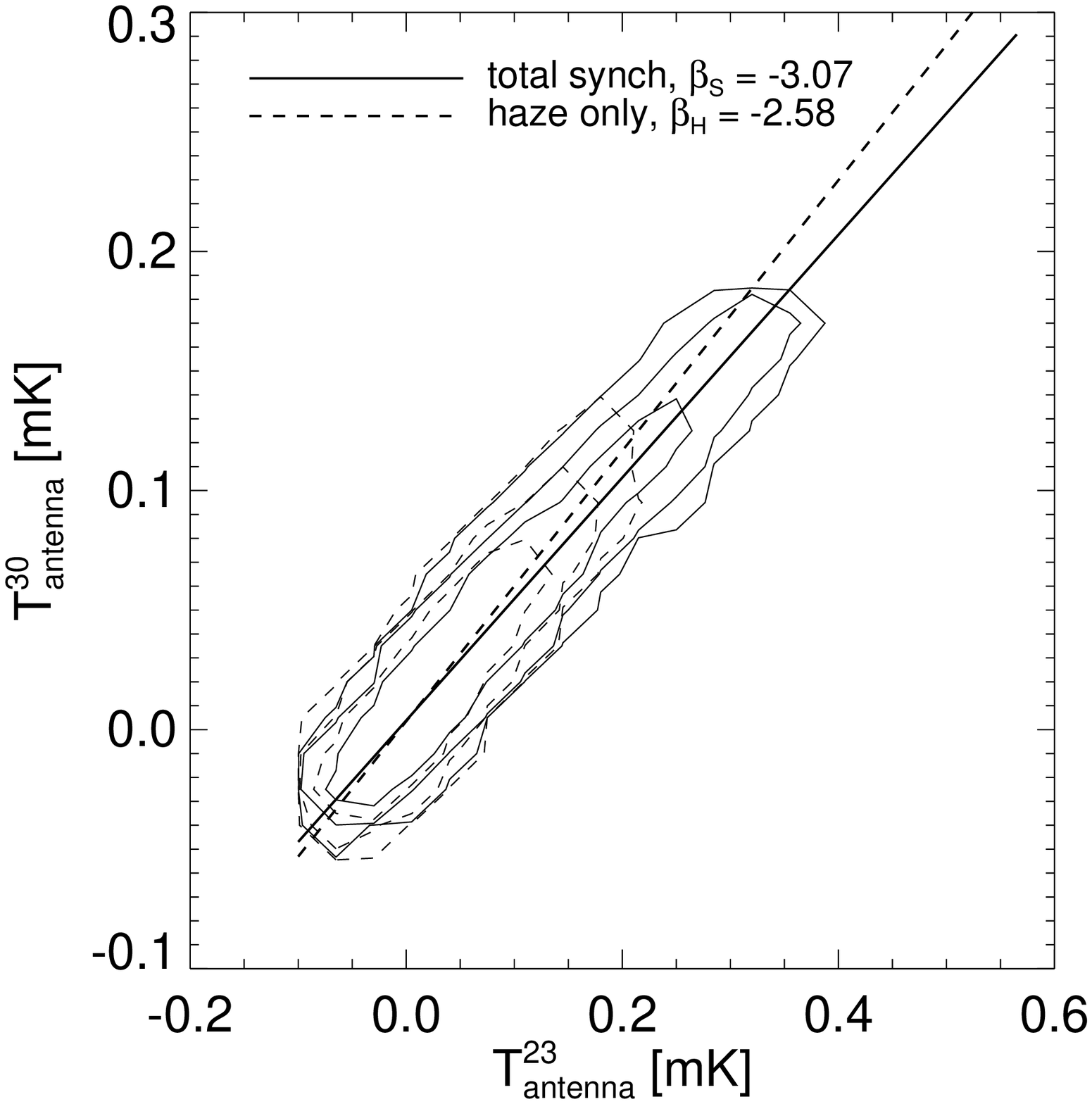}
    \includegraphics[width=0.33\textwidth]{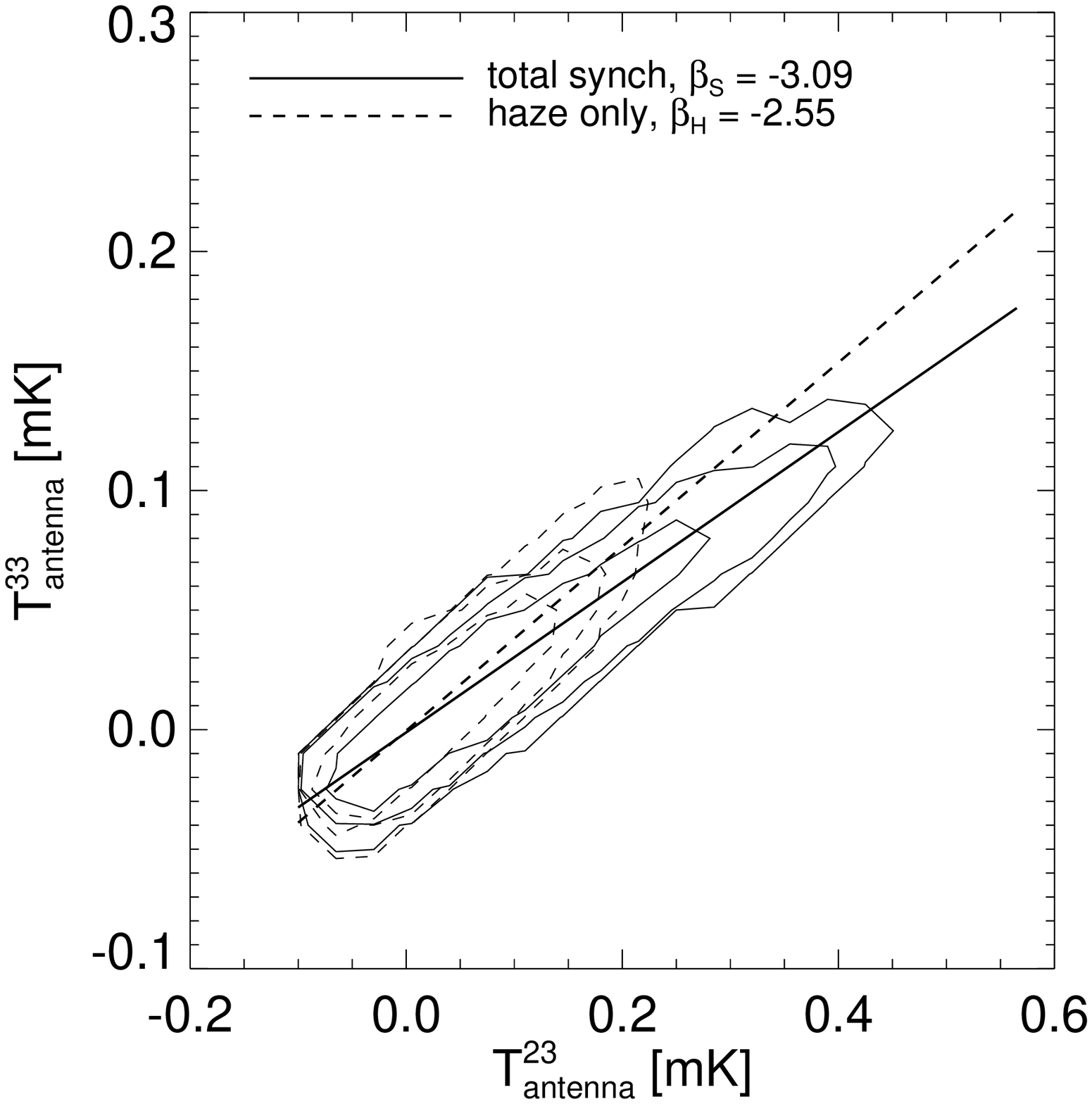}
  }
\caption{
  \emph{Left}: The spectrum measured from the residual
  in \reffig{rg-haze-all} in the region $|l| < 25\degr$,   $-35\degr < b < -10\degr$.  
  The haze spectrum is very nearly a power law with spectral index $\beta_H =
  -2.55$, while the total synchrotron emission in the region has a spectral
  index of $\beta_S = -3.1$ (see \refsec{specandmorph}),
  significantly \emph{softer} than the haze emission.  
  This spectrum should be free
  from biases due to template uncertainties.
  \emph{Middle} and \emph{right}: Scatter plots (shown in contours) for both the haze
  (dotted) and total synchrotron (solid) emission using \textit{WMAP} 23--33\,GHz
  and \textit{Planck} 30\,GHz.  }
\label{fig:haze-spec}
\epm

While a pixel-by-pixel determination of the haze spectrum is not
possible given the relatively low signal-to-noise ratio per pixel of
the haze emission, we can get a reliable estimate of its mean behaviour 
from the template fitting residuals in \reffig{rg-haze-all}.
The majority of previous haze studies have estimated the haze spectrum
via the template coefficients $\vec{a_{\nu}}$ for the haze
template. However, as noted in \citet{dobler12}, such an estimate
is not only affected by the CMB bias (which we have
effectively minimised by using the PILC), but may also be biased by
the effect of imperfect template morphologies.  The argument is as
follows: consider a perfectly CMB-subtracted map which consists
of the true haze $\vec{h}^{\prime}$ plus another true foreground
component $\vec{f}^{\prime}$ which we are approximating by templates
$\vec{h}$ and $\vec{f}$ respectively.  Our template fit approach can be
written as
\be
  a_{\rm H} \vec{h} + a_{\rm F} \vec{f} = b_{\rm H} \vec{h}^{\prime} + b_{\rm F}  \vec{f}^{\prime},
\ee
where we are solving for $a_{\rm H}$ and $a_{F}$ while $b_{\rm H}$ and $b_{F}$
are the true amplitudes.  The $a_{\rm H}$ solution to this equation is
\be
  a_{\rm H} = b_{\rm H} \times \frac{\Gamma_{\vec{h} \vec{h}^{\prime}}
  - \Gamma_{\vec{f} \vec{h}^{\prime}}\Gamma_{\vec{h} \vec{f}}}{1
  - \Gamma_{\vec{f} \vec{h}} \Gamma_{\vec{h} \vec{f}}} +
  b_{\rm F} \times \frac{\Gamma_{\vec{h} \vec{f}^{\prime}} - \Gamma_{\vec{f}
   \vec{f}^{\prime}}\Gamma_{\vec{h} \vec{f}}}{1 - \Gamma_{\vec{f}
  \vec{h}} \Gamma_{\vec{h} \vec{f}}},
\ee
where, for example, $\Gamma_{\vec{h} \vec{f}^{\prime}} \equiv \langle
\vec{h} \vec{f}^{\prime} \rangle/\langle \vec{h}^2 \rangle$, and the
mean is over unmasked pixels.  Thus, if $h=h^{\prime}$ and
$f=f^{\prime}$ then $a_{\rm H} = b_{\rm H}$ and we recover the correct spectrum.
However, if $h \neq h^{\prime}$ then the spectrum is biased and if
$f \neq f^{\prime}$ it is biased and dependent upon the true spectrum
of the other foreground, $b_{\rm F}$.

We emphasise that this bias is dependent on the cross-correlation of
the true foregrounds with the templates (which is unknown) and
that we have assumed a perfectly clean CMB estimate (which is not possible
to create) and have not discussed the impact of striping or other survey
artefacts (which Figs.~\ref{fig:fs-haze-all}
and \ref{fig:rg-haze-all} show are present).  Given this, a
much more straightforward estimate of the haze spectrum is to measure
it directly from $\mathcal{R}_{\rm H}$ in a region that is
relatively devoid of artefacts or other emission.  We measure the
spectrum in the GC south region $|l| < 35\degr$ and $-35\degr <
b < 0\degr$ by performing a linear fit (slope and offset) over
unmasked pixels and convert the slope measurement to a power law given
the central frequencies of the \planck\ and \textit{WMAP} data
(see \reffig{haze-spec}).  Specifically, we fit
\be
  \mathcal{R}_{\rm H}^{23} = A_{\nu} \times \mathcal{R}_{\rm H}^{\nu} + B_{\nu}
\ee
over unmasked pixels in this region for $A_{\nu}$ and $B_{\nu}$, and
calculate the haze spectral index, $\beta_{\rm H}=\log(A_{\nu})/\log(\nu/23\mbox{
GHz})$, for each $\nu$.  This spectrum should now be very clean
and -- given our use of the PILC -- reasonably unbiased.

A measurement of the spectrum of the haze emission is shown
in \reffig{haze-spec}.  It is evident that the \textit{WMAP} and \planck\
bands are complementarily located in log-frequency space and the two experiments together
provide significantly more information than either one
alone.\footnote{The close log-frequency spacing of the \textit{WMAP} 94\,GHz
and \planck\ 100\,GHz channels has the significant
advantage that the CO ($J$=1$\rightarrow$0) line falls in the \planck\ 100\,GHz band
while it is outside the \textit{WMAP} 94\,GHz band.  This provides an excellent
estimate for the CO morphology.}  In the left panel we plot
$\langle\mathcal{R}_{\rm H}^{\nu}\rangle-B_{\nu}$ (where the mean is over
the unmasked pixels in the region given above and the errors are their
standard deviation). The haze spectrum is measured to be
$T_{\nu} \propto \nu^{\,\beta_{\rm H}}$ with $\beta_{\rm H} = -2.55 \pm 0.05$.  This
spectrum is a nearly perfect power law from 23 to 41\,GHz.  Furthermore,
if we form the total synchrotron residual,
\be
  \mathcal{R}_{\rm S} = \mathcal{R}_{\rm H} + a_{S} \cdot \vec{s},
\ee
where $s$ is the Haslam map, and measure its spectrum in the south GC,
we again recover a nearly perfect power law with $\beta_{\rm S} = -3.1$.
Our conclusion is that the haze, which is not consistent
with free-free emission, arises from synchrotron emission with a spectral
index that is harder than elsewhere in the Galaxy by $\beta_{\rm H}
- \beta_{\rm S} = 0.5$.  Within the haze region, this component represents
$\sim33$\% of the total synchrotron and 23\% of the total Galactic
emission at 23\,GHz (\textit{WMAP} K-band) while emissions correlated with  Haslam, H$\alpha$, and
FDS contribute 43\%, 4\%, and 30\% respectively.

The $\beta_{\rm H} = -2.55$ spectral index of the haze is strongly
indicative of synchrotron emission from a population of electrons with
a spectrum that is harder than elsewhere in the Galaxy.  The other
possible origins of the emission in this frequency range (namely,
free-free and spinning dust) are strongly disfavored for several
reasons.  First, the spinning dust mechanism is very unlikely since
there is no corresponding feature in thermal dust emission at HFI
frequencies.  While it is true that environment can have an impact on
both the grain size distribution and relative ratio of spinning to
thermal dust emission (thus making the FDS models an imperfect tracer
of spinning dust, e.g., \citealt{ysard11}), to generate a strong
spinning dust signal at LFI frequencies while not simultaneously producing a
thermal signal a highly contrived grain population would be required,
in which small grains survive but large grains are completely
destroyed.  Furthermore, the FDS thermal predictions yield very low
dust-correlated residuals (see \reffig{rg-haze-all}) indicating a
close correspondence between thermal and spinning-dust morphology.
Finally, this spectrum is significantly softer than free-free
emission, which has a characteristic spectral index $\approx -2.15$.
Since the H$\alpha$ to free-free ratio is temperature-dependent, the
possibility exists that the haze emission represents some mixture of
synchrotron and free-free without yielding a detectable H$\alpha$
signal.  However, in order to have a measured spectral index of
$\beta_{\rm H} \approx -2.5$ from 23 to 41\,GHz, free-free could only
represent 50\% of the emission if the synchrotron component had a
spectral index $\approx -3$.  Since such a steep spectral index is
ruled out by the lack of a strong haze signal at 408\,MHz, the
synchrotron emission must have a harder spectrum and the free-free
component (if it exists) must be subdominant.\footnote{In addition,
  the lack of a bremsstrahlung signal in X-rays requires a fine tuning
  of the gas temperature to be $\sim10^6$\,K, a temperature at which
  the gas has a very short cooling time.  This also argues against a
  free-free explanation as described in \citet{mcquinn11}.} These
considerations, coupled with the likely inverse-Compton signal with
\textit{Fermi} \citep[see][]{dobler10,su10}, strongly indicate a
separate component of synchrotron emission.

\subsection{Spatial correspondence with the \textit{Fermi} haze/bubbles}
\label{sec:morphfermi}
\bpm
  \centerline{
    \includegraphics[width=0.9\textwidth]{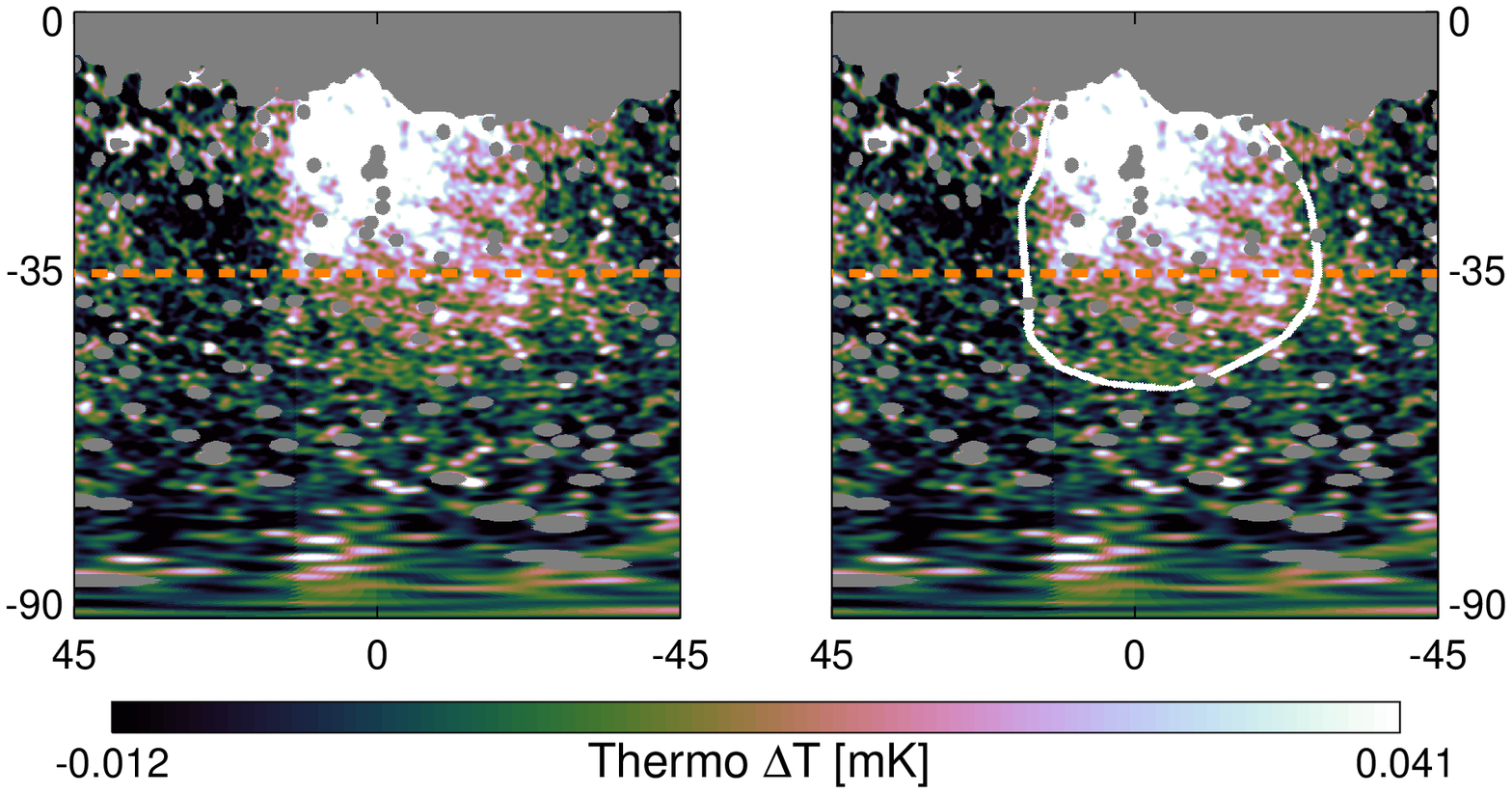}
    }
\caption{
  \emph{Left}: The southern \planck\ 30\,GHz haze
  from \reffig{rg-haze-all}.  \emph{Right}: The same but with contours of
  the \textit{Fermi} gamma-ray haze/bubbles \citep{su10} overlaid in white.  Above $b = -35\degr$
  (orange dashed line), the morphological correspondence is very
  strong suggesting that the two signals are generated by the same
  underlying phenomenon.
}
\label{fig:lowlatmorph}
\epm

\bpm
  \centerline{
    \includegraphics[width=0.9\textwidth]{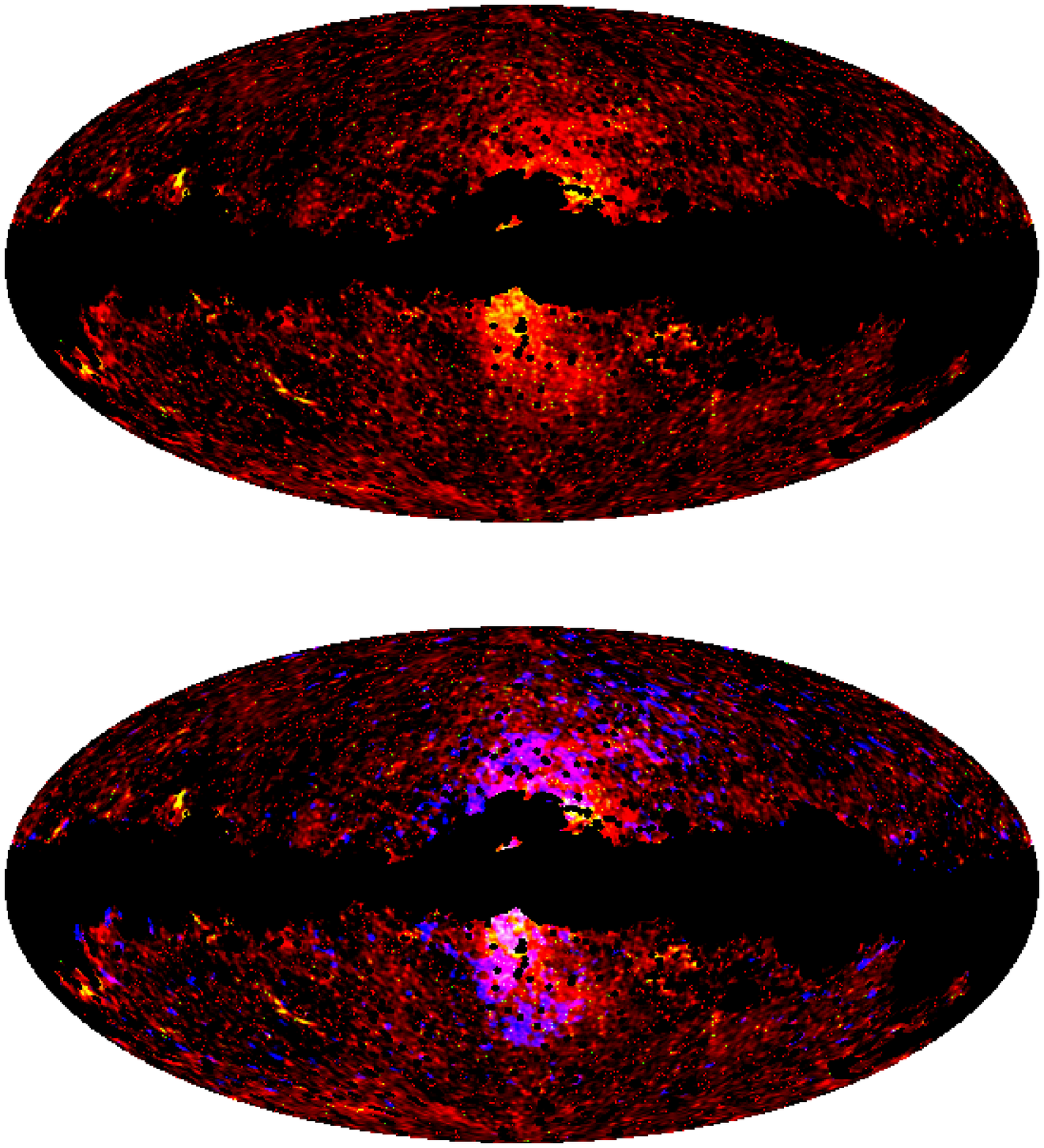}
    }
\caption{
  \emph{Top}: The microwave haze at \planck\ 30\,GHz (red, $-12$ $\mu$K
  $< \Delta T_{\rm CMB}$ $<$ 30 $\mu$K) and 44\,GHz (yellow, $12$ $\mu$K
  $< \Delta T_{\rm CMB}$ $<$ 40 $\mu$K).  \emph{Bottom}: The same but
  including the \textit{Fermi} 2-5 GeV haze/bubbles of \citet{dobler10}
  (blue, $1.05 <$ intensity $[$keV cm$^{-2}$ s$^{-1}$ sr$^{-1}] <
  1.25$; see their Fig.~11).  The spatial correspondence between the
  two is excellent, particularly at low southern Galactic latitude,
  suggesting that this is a multi-wavelength view of the same
  underlying physical mechanism.
}
\label{fig:rgb-planck}
\epm

The gamma-ray emission from the \textit{Fermi} haze/bubbles \citep{dobler10,su10}
is consistent with the inverse-Compton emission from a population of
electrons with the energy spectrum required to reproduce
the $\beta_{\rm H} = 2.55$ haze emission measured in this paper.
Furthermore, the \textit{Fermi} ``haze'' 
has a very strong spatial
coincidence with the \planck\ microwaves at low latitude (below
$|b|\sim35\degr$) as we show in \reffig{lowlatmorph}.  This suggests
a common physical origin for these two measurements with the gamma-ray
contribution 
extending down to $b\approx-50\degr$, while the microwaves fall off
quickly below $b\approx-35\degr$.  As in \citet{dobler12}, the
interpretation is that the magnetic field within the haze/bubbles
sharply decreases above $\sim5$ kpc from the Galactic plane while
the cosmic-ray distribution extends to $\sim10$ kpc and continues
to generate gamma-ray emission (e.g., by inverse Compton scattering CMB
photons).  In \reffig{rgb-planck} we show a full-sky 
representation of the \planck\ haze emission overlaid with
the \textit{Fermi} gamma-ray haze/bubbles from \citet{dobler10}.

\section{Summary}
\label{sec:summary}
We have identified the presence of a microwave haze in the \textit{Planck} LFI
data and performed a joint analysis with 7-year \textit{WMAP} data.  Our
findings verify not only that the haze is real, but also that it is
consistent in amplitude and spectrum in these two different
experiments.  Furthermore, we have used \textit{Planck} HFI maps to
generate a CMB estimate that is nearly completely clean of haze
emission, implying that we have reduced systematic biases in the
inferred spectrum to a negligible level.  We find that the unbiased
haze spectrum is consistent with a power law of spectral index
$\beta_{\rm H} = -2.55 \pm 0.05$, ruling out free-free emission as a possible
explanation, and strengthening the possibility of a hard synchrotron
component origin. The
spectrum of softer synchrotron emission found elsewhere in the Galaxy is $\beta_{\rm S} =
-3.1$, consistent with a cosmic-ray electron population that has been
accelerated in supernova shocks and diffused throughout the Galaxy.
This spectrum is significantly softer than the haze emission,
which is not consistent with supernova shock acceleration after taking
into account energy losses from diffusion effects.

The microwave haze is detected in the \planck\ maps with both simple
template regression against the data and a more sophisticated Gibbs
sampling analysis.  The former provides an excellent visualisation of
the haze at each wavelength on large scales while the latter allows
a pixel-by-pixel analysis of the complete data set.
While the template analysis allows us to derive the
$\beta_{\rm H} = -2.55$ spectrum with high confidence, 
spectral determination with the Gibbs approach is more difficult
given that noise must be added to the analysis to ensure 
convergence in the sampling method, and that a significantly 
more flexible model (in particular, one in which the
spectrum of synchrotron is allowed to vary with each pixel) is used. 
However, not only
is the spatial correspondence of the haze derived with the two methods
excellent, but the Gibbs method allows us to show conclusively that
the microwave haze is a separate component and not merely a variation in
the spectral index of the synchrotron emission.

The morphology of the microwave haze is nearly identical from
23 to 44\,GHz, implying that the spectrum does not
vary significantly with position.  Although detection of the haze in
polarisation with \textit{WMAP} remains unlikely given the noise level of the data
\citep{dobler12}, future work with \textit{Planck} will concentrate on 
using its enhanced sensitivity to search for this component.

\begin{acknowledgements}
The development of \Planck\ has been supported
by: ESA; CNES and CNRS/INSU-IN2P3-INP (France); ASI, CNR, and INAF
(Italy); NASA and DoE (USA); STFC and UKSA (UK); CSIC, MICINN and JA
(Spain); Tekes, AoF and CSC (Finland); DLR and MPG (Germany); CSA
(Canada); DTU Space (Denmark); SER/SSO (Switzerland); RCN (Norway);
SFI (Ireland); FCT/MCTES (Portugal); and DEISA (EU). A description of
the Planck Collaboration and a list of its members, including the
technical or scientific activities in which they have been involved,
can be found at \url{http://www.rssd.esa.int/Planck}.
G. Dobler has been supported by the Harvey L. Karp Discovery Award.
Some of the results in this paper have been derived using the HEALPix \citep{gorski05} package.
\end{acknowledgements}

\bibliographystyle{aa}
\bibliography{planck_haze,Planck_bib}

\raggedright  
\end{document}